\documentclass[twocolumn]{aastex63}
\usepackage{ amsmath}

\defcitealias{vanVelzen+21}{V21}

\newcommand{\be}{\begin{equation}}
\newcommand{\ee}{\end{equation}}

\shorttitle{A Unified Theory of Jetted Tidal Disruption Events}
\shortauthors{O.~Teboul \& B.~D.~Metzger}
\graphicspath{{./}{figures/}}

\begin{document}

\title{A Unified Theory of Jetted Tidal Disruption Events:\\
From Promptly Escaping Relativistic to Delayed Transrelativistic Jets}

\correspondingauthor{Odelia Teboul}
\email{odelia.teboul1@mail.huji.ac.il}

\author{Odelia Teboul}
\affil{Racah Institute of Physics, The Hebrew University, 91904, Jerusalem, Israel}

\author[0000-0002-4670-7509]{Brian D.~Metzger}
\affil{Department of Physics and Columbia Astrophysics Laboratory, Columbia University, New York, NY 10027, USA}
\affil{Center for Computational Astrophysics, Flatiron Institute, 162 5th Ave, New York, NY 10010, USA} 

\begin {abstract}
Only a tiny fraction $\sim 1\%$ of stellar tidal disruption events (TDE) generate powerful relativistic jets evidenced by luminous hard X-ray and radio emissions. We propose that a key property responsible for both this surprisingly low rate and a variety of other observations is the typically large misalignment $\psi$ between the orbital plane of the star and the spin axis of the supermassive black hole (SMBH). Such misaligned disk/jet systems undergo Lense-Thirring precession together about the SMBH spin axis. We find that TDE disks precess sufficiently rapidly that winds from the accretion disk will encase the system on large scales in a quasi-spherical outflow. We derive the critical jet efficiency $\eta > \eta_{\rm crit}$ for both aligned and misaligned precessing jets to successfully escape from the disk-wind ejecta. As $\eta_{\rm crit}$ is higher for precessing jets, less powerful jets only escape after alignment with the SMBH spin. Alignment can occur through magneto-spin or hydrodynamic mechanisms, which we estimate occur on typical timescales of weeks and years, respectively. The dominant mechanism depends on $\eta$ and the orbital penetration factor $\beta$. Hence depending only on intrinsic parameters of the event $\{\psi,\eta,\beta\}$, we propose that each TDE jet can either escape prior to alignment, thus exhibiting erratic X-ray light curve and two-component radio afterglow (e.g., Swift J1644+57) or escape after alignment. Relatively rapid magneto-spin alignments produce relativistic jets exhibiting X-ray power-law decay and bright afterglows (e.g., AT2022cmc), while long hydrodynamic alignments give rise to late jet escape and delayed radio flares (e.g., AT2018hyz).
\end {abstract}

\section{Introduction}
\label{sec:intro}
After 11 years of hibernation, another jetted tidal disruption event (TDE) was recently observed in the form of the optically-discovered AT 2022cmc \citep{Andreoni22}, increasing their number to the strikingly low total of 4.  The three previous on-axis jetted TDEs were discovered via $\gamma$-ray triggers by the Neil Gehrels Swift Observatory:  Swift J1644+57 (hereafter J1644; \citealt{Bloom+11,Burrows+11,Levan+11}), Swift J2058.4+0516 (J2058; \citealt{Cenko12}) and Swift J1112.28238 (J1112; \citealt{Brown15, Kawamuro16}).  While J1644 exhibited highly variable hard X-ray emission for the first $\sim 10$ days (seemingly distinct from the other events), all of these promptly jetted TDE exhibited late-time X-ray emission following a power-law decay $L_{\rm X} \propto t^{-\alpha}$ with $\alpha \approx 5/3 - 2.2$, broadly similar to the expected mass fall-back rate following a full or partial stellar disruption (e.g., \citealt{Rees88,Guillochon&RamirezRuiz13}).  The X-ray jetted TDE were also accompanied by bright synchrotron radio emission lasting for years \citep{Berger12, Zauderer+13, Cenko12, Brown17}, powered by the shock interaction of the jet material with the surrounding circumnuclear medium \citep{Giannios&Metzger11}.  In addition to these conspicuously jetted events, a small but growing sample of optically-selected TDE show radio afterglows consistent with non relativistic outflows (e.g., \citealt{Alexander16, vanVelzen16,Alexander+17,Saxton17,Mattila18}; see \citealt{Alexander+20} for a review).  Moreover, the radio emission in some cases is seen to rise beginning only many months to years after discovery  (\citealt{Horesh+21,Horesh+21b,Perlman+22,Cendes+22,Sfaradi+22}), pointing to outflows or jets which only become observable after a significant delay from the time of disruption and the peak of the optical flare. 

The stellar debris that remains bound to the supermassive black hole (SMBH) following the disruption should eventually form an accretion disk around the SMBH near the tidal radius at tens to hundreds of gravitational radii (e.g., \citealt{Rees88}).  The timescale and mechanism of disk accretion depend on the relative efficiency of three processes: circularization, radiative cooling, and viscous accretion (e.g., \citealt{Evans&Kochanek89,Metzger22,Bonnerot&Stone21}).  Even once a disk forms, the accreting material cannot cool efficiently because radiation is trapped in the highly super-Eddington accretion flow, leading to powerful disk winds (e.g., \citealt{Strubbe&Quataert09,Ohsuga&Mineshige11,Coughlin&Begelman14,Metzger&Stone16,Dai+18,Dai+21}) and relativistic jets (e.g., \citealt{Giannios&Metzger11,Sadowski&Narayan15,Sadowski&Narayan16,Curd&Narayan19,Coughlin&Begelman20}).  In both J1644 and J2058, the jetted X-ray emission abruptly terminated on timescale of years, around the expected transition time from super-Eddington to sub-Eddington fall-back \citep{Zauderer+13,Pasham+15}, supporting super-Eddington accretion rates as a necessary condition for powerful early jet launching \citep{Tchekhovskoy+14}. 

The inferred volumetric rate of on-axis jetted TDEs ($\approx 0.03 \: {\rm Gpc}^{-3} {\rm yr}^{-1}$; e.g., \citealt{Burrows+11,Andreoni22}), represents only a tiny fraction $f_{\rm on} \sim 10^{-4} - 10^{-3} $ of the total TDE rate ($\sim 10^{2}- 10^{3} $ Gpc$^{-3}$ yr$^{-1}$; e.g., \citealt{Odelia2,Yao+23}).  This discrepancy can in part be attributed to geometric and relativistic beaming of the jetted X-ray/gamma-ray emission.  However, adopting values for the Lorentz factor of the (unshocked) jet $\Gamma_{\rm j} \sim 10$ and half-opening angle $\theta_{\rm j} \sim 1/\Gamma_{\rm j} \sim 0.1$ typical of blazar jets and those found by modeling the radio afterglow of J1664 (e.g., \citealt{Metzger+12}) and AT 2022cmc (e.g., \citealt{Matsumoto&Metzger23}), the corresponding emission beaming fraction $f_{\rm b} \simeq \theta_{\rm j}^{2}/2 \sim 10^{-2}$ is still orders of magnitude higher than $f_{\rm on}$.  This points to an intrinsic prompt jetted fraction of only $f_{\rm on}/f_{\rm b} \sim 1\%$.  The lack of bright off-axis radio emission from large samples of optical- and X-ray-selected TDEs also points to a small promptly-jetted fraction (e.g., \citealt{Bower+13,vanVelzen+13,Generozov+17}; however, see \citealt{Somalwar+23} for a candidate orphan radio TDE).  

Why would only a small fraction of TDEs launch (initially) successful jets?  One possible explanation is that jet production through the commonly invoked \citet{Blandford&Znajek77} mechanism requires special and rare conditions, such as a high SMBH spin and/or a strong magnetic flux threading the disrupted star \citep{Giannios&Metzger11,Tchekhovskoy+14}. However, assuming that all TDE jets carry similar magnetic flux and that the spin distribution of quiescent SMBHs is similar to those in Active Galactic Nuclei (AGN; e.g., \citealt{Reynolds13}), promptly jetted TDEs should be much more common than the $\sim 1\%$ observed. 

Here, we propose that a key factor responsible for both the low rate of prompt jetted TDEs and the growing sample of TDEs that exhibit delayed radio jet emission is the distribution of initial misalignment angles between the equatorial plane of the SMBH and the orbital plane of the disrupted stars.  Indeed, the orbital angular momentum vector of the disrupted star in a TDE should be drawn from a roughly isotropic distribution, rendering the probability for it to coincide with the spin axis of the SMBH to within the solid angle of a jet $\lesssim f_{\rm b}$ to be commensurably small $\sim 1 \%$.  Even if all TDEs create jets, {\it the vast majority will thus (at least initially) be misaligned with the SMBH spin}.  

\begin{figure*}
    \centering
    \includegraphics[width=1.0\textwidth]{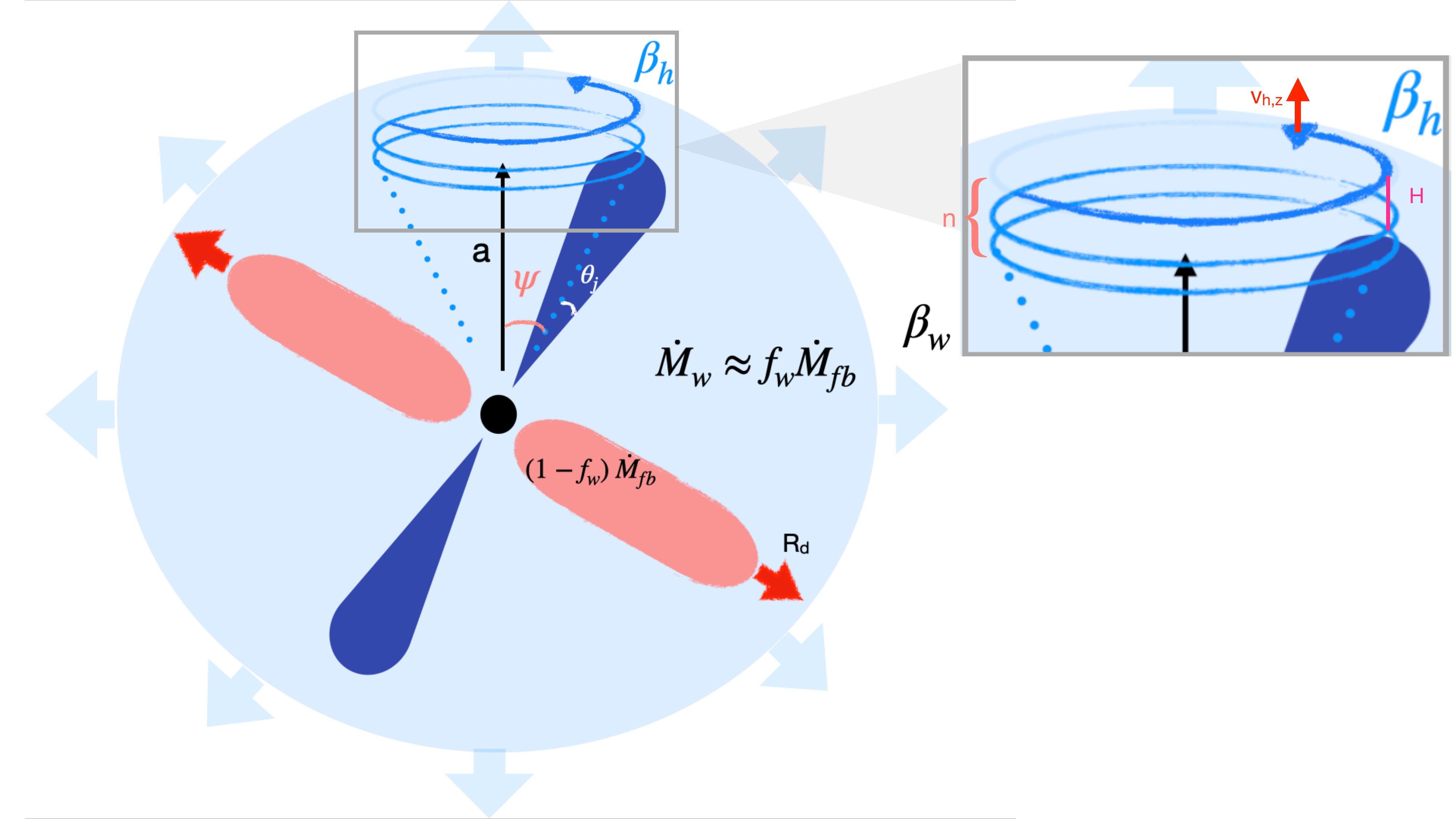}
    \caption{Schematic illustration of a misaligned jetted TDE. Stellar debris following the disruption return to the orbital pericenter and form an accretion disk in the orbital plane of the star whose normal is in general significantly misaligned by an angle $\psi$ with respect to the SMBH spin axis.  Matter falls back to the disk at a highly super-Eddington accretion rate $\dot{M}_{\rm fb}$, driving a disk-wind of velocity $\beta_{\rm w} = v_{\rm w}/c$ and mass-loss rate $\dot{M}_{\rm w} \approx f_{\rm w} \dot{M}_{\rm fb}$.  The remainder of the fall-back feeds the SMBH at a rate $\dot{M} = (1-f_{\rm w})\dot{M}_{\rm fb}$, powering a jet of luminosity $L_{\rm j} = \eta \dot{M}c^{2}$ and half-opening angle $\theta_{\rm j}$. Due to rapid precession of the disk orientation, on sufficiently large scales $(R \gtrsim R_{\rm min}$; Eq.~\eqref{eq:Rmin}) the wind will collide and mix with itself and generate an approximately spherical outflow.  The jet precesses with the disk and its head propagates into the disk outflow with velocity $\beta_{\rm h}$; precession effectively spreads the jet energy across a wider cone of half-opening angle $\psi$, the disk/spin misalignment angle. The insert shows a zoomed-in image of quantities relevant to the propagation of the helical jet (see Sec.~\ref{sec:propagation} for details). }
    \label{fig:cartoon}
\end{figure*}

Tilted thick accretion disks undergo solid body precession about the SMBH spin axis as a result of Lense-Thirring torques (e.g., \citealt{Fragile+07,Stone&Loeb12,Franchini+16}). Moreover, GRMHD simulations of accreting black holes show that$-$due to collimation provided by mass-loaded wide-angle disk outflows$-$the jet propagates on large scales along the disk rotation axis, precessing together with the disk (e.g., \citealt{Liska+18,Chatterjee20}).  Sufficiently rapid disk precession will thus result in the disk outflows which emerge along any direction colliding on large radial scales with outflows released along the same direction from earlier phases of the precession cycle.  This interaction will induce mixing between the wind material, encasing the jet in an environment that we shall assume can on large scales be roughly approximated as a spherical outflow or envelope (see Fig. \ref{fig:cartoon} for a schematic illustration). {\it This implies that rapidly-precessing jets need to escape from the disk-wind ejecta in order to be successful.}

Although the disk/jet system is initially misaligned with the SMBH spin, various mechanisms can lead to alignment with time following the TDE (e.g., \citealt{Papaloizou&Pringle83,McKinney+13,Franchini+16}). These mechanisms can be broadly separated into two categories: i) ``hydrodynamic" alignment mechanisms, which result from internal stresses within the accretion disk itself, and ii) ``electromagnetic" mechanisms, which result from large-scale magnetic stresses on the disk by the jet. In this work, we investigate the conditions necessary for TDE jets to be successful for different types of disk-spin alignment, and explore the resulting observational consequences.

This paper is organized as follows. In Section \ref{sec:essential} we review some essential elements of TDEs and SMBH jet formation.  In Section \ref{sec:disk/jet}, we investigate the evolution of misaligned TDEs disk taking into account viscous spreading and wind launching. We then derive the disk/jet precession and alignment timescales for the different relevant mechanisms.  In Section \ref{sec:propagation}, we derive the impact of precession on the disk-wind and the conditions for both aligned and misaligned precessing jets to successfully escape from the disk-wind.  In Section \ref{sec:escape} we propose a unified model that can explain both promptly escaping jet and delayed radio emission depending only on the intrinsic TDE parameters. Then, for each class, we compute typical energy, X-ray and radio light curves. Finally, for each class, we compute the radio observations of an exemplar TDE.  

\section{Tidal Disruption Event and Jet Formation}
\label{sec:essential}
\subsection{Tidal Disruption Event}
\label{sec:set-up}

A Sun-like star of mass $M_{\star} = M_{\odot}$ and radius $R_{\star} = R_{\odot}$ that wanders too close to a SMBH of mass $M_{\bullet} = 10^{6}M_{\bullet,6}M_{\odot}$ will be tidally disrupted if the pericenter radius of its orbit, $R_{\mathrm{p}}$, becomes less than the tidal radius,
\begin{equation}
R_{\mathrm{t}}  = R_{\star}\left(M_{\bullet}/M_{\star}\right)^{1/3} \approx 7 \times 10^{12}  M_{\bullet, 6}^{1/3} \mathrm{~cm} \approx 47 M_{\bullet, 6}^{-2/3} R_{\mathrm{g}},
\label{eq:Rt}
\end{equation}
where $R_{\mathrm{g}} \equiv G M_{\bullet} / c^2$.  Defining the orbital penetration factor as $\beta \equiv R_{\mathrm{t}} / R_{\mathrm{p}} \geq 1$, the star will be disrupted outside of the horizon for
\be
\beta \lesssim \beta_{\rm max} \simeq 12 M_{\bullet,6}^{-2/3}(R_{\rm ibso}/4R_{\rm g})^{-1},
\label{eq:betamax}
\ee
where $R_{\rm ibso}$ is the radius of the innermost bound spherical orbit, normalized to $\simeq 4R_{\rm g}$ for a slowly-spinning SMBH.

The disruption process binds roughly half the star to the SMBH by a specific energy $\left|E_{\mathrm{t}}\right| \sim G M_{\bullet} R_{\star} / R_{\mathrm{t}}^2$ corresponding roughly to the difference in the tidal potential over the stellar radius. The most tightly bound matter falls back to the SMBH on the characteristic timescale set by the orbital period corresponding to this binding energy:
\begin{equation}
t_{\mathrm{fb}} \simeq 41  M_{\bullet, 6}^{1/2}  \mathrm{~d}.
\label{eq:tfb}
\end{equation}

We shall focus on complete disruptions, for which the resulting mass fall-back rate at times $t \gg t_{\mathrm{fb}}$ obeys (e.g., \citealt{Rees88,Phinney89,Lodato+09}),
\begin{equation}
\dot{M}_{\mathrm{fb}} = \dot{M}_{\rm p}\left(\frac{t}{t_{\mathrm{fb}}}\right)^{-5/3},
\label{eq:Mdotfb}
\end{equation}
where the peak fall-back rate
\begin{equation}
\dot{M}_{\rm p} = \frac{M_\star}{3 t_{\mathrm{fb}}} \approx 1.9 \times 10^{26} M_{\bullet,6}^{-1/2} \mathrm{~g} \mathrm{~s}^{-1},
\end{equation}
can exceed the Eddington accretion rate $\dot{M}_{\rm Edd} \equiv L_{\rm Edd}/0.1 c^{2} \approx 1.7\times 10^{24}$ g s$^{-1}$ by orders of magnitude, with $L_{\rm Edd} \simeq 1.5\times 10^{44}M_{\bullet,6}$ erg s$^{-1}$.  The disk will become sub-Eddington ($\dot{M}_{\rm fb} = \dot{M}_{\rm Edd}$) after a time
\be
t_{\rm Edd} \simeq 760 \,M_{\bullet,6}^{-2/5}\,\,{\rm d}.
\label{eq:tEdd}
\ee
Although the details and timescales of the debris circularization process remain areas of active research (e.g., \citealt{Bonnerot&Stone21,Metzger22}), the returning debris material eventually collides with itself, dissipating its bulk kinetic energy. For jetted TDE, the circularization likely occurs quickly. Hence hereafter we assume that within $\lesssim$ few $t_{\rm fb}$, the debris settles into rotationally-supported disk near the circularization radius 
\be R_{\rm c} = 2R_p/\beta \simeq 1.4\times 10^{13} \beta^{-1}M_{\bullet,6}^{1/3} \mathrm{~cm},
\label{eq:Rc}
\ee
which subsequently feeds the SMBH at a rate $\dot{M} \sim \dot{M}_{\rm fb}$. 

\subsection{Jets and Outflows}
\label{sec:jet}

Radiation is trapped in the accretion flow on the inflow time out to the ``trapping'' radius \citep{Begelman79}
\be
R_{\rm Edd} \simeq 10 R_{\rm g}\left(\frac{\dot{M}}{\dot{M}_{\rm Edd}}\right).
\label{eq:Redd}
\ee
Because initially $R_{\rm Edd} \gg R_{\rm c}$, the accreting gas cannot cool radiatively, rendering the geometrically-thick accretion flow marginally bound and subject to powerful disk outflows (e.g., \citealt{Strubbe&Quataert09,Ohsuga&Mineshige11,Coughlin&Begelman14,Metzger&Stone16,Dai+18,Dai+21}) and relativistic jets (e.g., \citealt{Giannios&Metzger11,Sadowski&Narayan15,Sadowski&Narayan16,Curd&Narayan19,Coughlin&Begelman20}).  
We assume that the SMBH feeds a relativistic jet with a time-dependent kinetic luminosity that we parameterize according to
\be 
L_{\rm j}(t) \simeq \eta \dot{M}(t) c^{2} \propto \eta t^{-5/3}, 
\label{eq:Ljet}
\ee
where $\eta$ is a potentially time-dependent dimensionless efficiency parameter.  In what follows we shall consider $\eta$ to be a free parameter, insofar that its magnitude and time evolution depend on the physical mechanism powering the jet (e.g., radiation- versus MHD-powered mechanisms) and uncertain properties such as the SMBH spin.

A common mechanism invoked to power black hole jets is via the \citet{Blandford&Znajek77} (BZ) process, in which a strong magnetic field threading the black hole extracts its spin energy via a large-scale Poynting-flux-dominated outflow.  GRMHD simulations find that the BZ jet efficiency is given by (e.g., \citealt{Tchekhovskoy+10})
\begin{equation}
\eta_{\rm BZ} \simeq \frac{\kappa}{16\pi R_{\rm g}^{2}}\frac{\Phi_{\bullet}^{2}}{\dot{M} c}\omega_{\rm H}^{2}f(\omega_{\rm H}) \simeq 2\left(\frac{\Upsilon}{10}\right)^{2}\omega_{\rm H}^{2}f(\omega_{\rm H}),
\label{eq:etaBZ}
\end{equation}
where $\Phi_{\bullet}$ is the magnetic flux threading the horizon,
$\kappa \simeq 0.045$, $\omega_{\rm H}= a/(1+(1-a^{2})^{1/2})$ is the dimensionless angular frequency of the SMBH, $a$ is the dimensionless SMBH spin, and $f(\omega_{\rm H}) \simeq 1+0.35 \omega_{\rm H}^{2}-0.58 \omega_{\rm H}^{4}$ is a high-spin correction.  In the final line we have introduced the dimensionless magnetic flux,
\be
\Upsilon \equiv \frac{0.7\Phi_{\bullet}}{\sqrt{4\pi R_{\rm g}^{2}\dot{M}c}},
\label{eq:Upsilon}
\ee
such that $\Upsilon \gtrsim \Upsilon_{\rm max} \approx 10-20$ corresponds to a ``magnetically-arrested'' (MAD) disk \citep{Narayan+03,Tchekhovskoy+10}.  In the MAD limit, $\eta$ saturates to a maximum value varying from $\eta_{\rm max} \approx 0.3$ (for $a \approx 0.5$) to $\eta_{\rm max} \approx 1.4$ (for $a = 0.99$; \citealt{Tchekhovskoy+11}).  Although the magnetic flux threading the SMBH in a TDE is highly uncertain (e.g., \citealt{Giannios&Metzger11,Kelley+14}), a high jet efficiency $\eta \sim \eta_{\rm max} \sim 1$ is likely necessary to explain extremely powerful jets such as in J1644 \citep{Tchekhovskoy+14}.

\section{Disk/Jet Precession and Alignment}
\label{sec:disk/jet}
As discussed in Sec. \ref{sec:intro}, the orbital plane of the star will in general be misaligned with the SMBH spin axis in the vast majority of TDEs.  Such tilted thick disks undergo Lense-Thirring precession around the SMBH spin axis (e.g., \citealt{Fragile+07,Stone&Loeb12,Franchini+16}), with the jet precessing together with the disk (e.g., \citealt{Liska+18}).
In this section we explore the time evolution of the misaligned disk (Sec.~\ref{sec:disk}) and compute its precession rate around the BH spin axis for different configurations (Sec.~\ref{sec:precession}), before estimating the conditions and timescales for the different disk/jet alignment mechanisms (Sec.~\ref{sec:alignment}).

\subsection{Disk Evolution}
\label{sec:disk}
After a few orbits, the returning bound debris material collides with itself settling at its circularization radius (Eq. \ref{eq:Rc}), before accreting onto the SMBH. The fall-back material which returns to the SMBH feeds a gaseous disk of radial surface density $\Sigma(r,t)$.  The latter evolves under the action of viscosity according to the standard diffusion equation \citep{Pringle81} 
\begin{equation}
\frac{\partial \Sigma}{\partial t}-\frac{1}{r} \frac{\partial}{\partial r}\left[3 r^{1 / 2} \frac{\partial}{\partial r}\left(\Sigma \nu r^{1 / 2}\right)\right] = (1-f_{\rm w}) \frac{\dot{M}_{\mathrm{fb}}(t)}{2 \pi R_{\mathrm{c}}}\delta\left(r-R_{\mathrm{c}}\right),
\label{eq:Sigma}
\end{equation}
where the source-term on the right-hand side accounts for the mass added to the disk by fall-back accretion near the circularization radius $R_{\rm c}$ (Eq.~\eqref{eq:Rc}) at the rate $\dot{M}_{\rm fb}(t)$ (Eq.~\ref{eq:Mdotfb}), as expected for a misaligned disk  (\citealt{Shen&Matzner14}), and the  $1-f_{\rm w}$ factor accounts for the fraction $f_{\rm w}$ lost to disk winds (see Fig.~\ref{fig:cartoon}, Sec.~\ref{sec:wind}).  The accretion disk is expected to be radially-narrow, geometrically thick (e.g., \citealt{Strubbe&Quataert09}) and advective at least until the transition to the radiative state \citep{Shen&Matzner14}. The inner radius of the disk, $R_{\rm in}$ corresponds to the innermost stable circular orbit (ISCO) for an aligned disc and to the innermost stable spherical orbit (ISSO) for a misaligned disk (e.g., \citealt{Stone+13}).  In the top left panel of Fig.~\ref{fig:Tprec} we show the inner radius for a SMBH of mass $10^6 M_{\odot}$ for different spins $a$ spanning both prograde and retrograde values for a few different misalignment angles $\psi = 30^{\circ}, 60^{\circ}$ in comparison to the ISCO radius.

We adopt a kinematic viscosity of the standard parameterized form \citep{Shakura&Sunyaev73} 
\begin{equation}
\nu=\alpha r^2 \Omega\left(\frac{H}{r}\right)^2=\nu_{\mathrm{c}}\left(\frac{r}{R_{\mathrm{c}}}\right)^{1/2},
\label{eq:nu}
\end{equation}
where $\alpha$ is the dimensionless viscosity parameter, $\Omega=(G M_{\bullet} / r^3)^{1/2}$ and $H$ is the vertical scale-height.  In the final line we have taken $H/r \sim \mathcal{O}(1) $ constant which is appropriate for a thick disk and defined $\nu_{\mathrm{c}} \equiv \nu\left(R_{\mathrm{c}}\right)$.  

The viscous time at the circularization radius,
\be
t_{\nu, \rm c} =\frac{R_{\rm c}^2}{\nu_{\rm c}} \simeq 5.8  \beta^{-3/2}\alpha_{-1}^{-1}\left(\frac{H}{0.3 r}\right)^{-2} \text{d},
\label{eq:tnuc}
\ee
where $\alpha_{-1} = \alpha/(0.1)$, is typically short compared to the fall-back time (Eq.~\eqref{eq:tfb}) as long as the disk remains in a thick super-Eddington or slim-disk state.  A radially-steady accretion flow $\dot{M}(r) \sim \dot{M}_{\rm fb}$ is established at small radii $r \lesssim R_{\rm c}$ extending down to the SMBH, while a smaller fraction of infalling gas spreads to larger radii outside $R_{\rm c}$ in order to carry away the deposited angular momentum.  In particular, for a power-law viscosity of the form \eqref{eq:nu}, the solution to Eq.~\eqref{eq:Sigma} when the source-term $\propto \dot{M}_{\rm fb}$ evolves slowly compared to the viscous time (as is at least initially satisfied if $t_{\nu,\rm c} \lesssim t_{\rm fb}$) can be found analytically using Green's function (\citealt{Metzger+12b}). Hence, the surface density is approximately given by 
\be
\Sigma(r,t) \simeq  \frac{1}{3 \pi} \frac{\dot{M}_{\mathrm{fb}} t_{\nu, \mathrm{c}}} {R_{\mathrm{c}}^2}
\begin{cases}
\left(\frac{r}{R_{\mathrm{c}}}\right)^{-1/2}, R_{\rm in} < r<R_{\mathrm{c}} \\
  \left(\frac{r}{R_{\mathrm{c}}}\right)^{-1}, R_{\mathrm{c}} < r < R_{\rm d},
    \end{cases}  
    \label{eq:Sigma_analytic}
\ee
where (e.g., \citealt{Metzger+08,Shen&Matzner14})
\be 
R_{\rm d} \approx R_{\rm c}\left(\frac{t}{t_{\nu, \rm c}}\right)^{\chi},
\label{eq:Rd}
\ee 
is the outer radius to which the disk has spread viscously in time $t$.  The spreading rate can vary from $\chi \in [2/5,2/3]$, depending on the efficiency of angular momentum losses due to disk outflows (see \citealt{Metzger+08}; their Eq.~B14 and surrounding discussion).  The spreading rate at large radii can also vary at late times $t \gg t_{\rm fb}$, once the local viscous ($\sim$ spreading) timescale exceeds the fall-back time, and the assumption used to derive Eq.~\eqref{eq:Sigma_analytic} breaks down.

\subsection{Disk/Jet Precession}
\label{sec:precession}

Analytic calculations \citep{Papaloizou&Pringle83,Papaloizou&Lin95}, Newtonian hydrodynamic simulations \citep{Larwood&Papaloizou97, Nelson&Papaloizou00}, and full GRMHD simulations (\citealt{Fragile+07,Dexter&Fragile11,Liska+18}) show that thick accretion disks with aspect ratios $H/r \gtrsim \alpha$ generally precess as solid bodies following Lense-Thirring precession.  The solid-body precession period for a thick disk extending from radii $r \in [R_{\rm in},R_{\rm out}]$ is given by $T_{\rm prec}=2 \pi \sin \psi (J/\mathcal{N})$, where
\begin{equation}
J (t) =2 \pi \int_{R_{\rm in}}^{R_{\rm out}(t)} \Sigma(r,t) \Omega(r) r^3 \mathrm{~d} r
\label{eq:J}
\end{equation}
is the total angular momentum of the disk and 
\begin{equation}
\mathcal{N} (t) =4 \pi \frac{G^2 M ^2 a}{c^3} \sin \psi \int_{R_{\rm in}}^{R_{\rm out}(t)}  \Sigma(r,t) \Omega(r)  \mathrm{~d} r,
\end{equation}
 is the radially-integrated Lense-Thirring torque.  Both quantities are time-dependant allowing us to take into account the spreading of the outer radius (Eq. \ref{eq:Rd}). For the broken power-law density profile Eq.~\eqref{eq:Sigma_analytic}, we find:
\begin{equation}
   T_{\rm prec}= \frac {\pi  c^3}{G^2 M_{\bullet}^2 a}\frac{\frac{2}{3} R_{\rm c}^{1/2}R_{\rm out}(t)^{3/2} -\frac{1}{6}R_{\rm c}^2 -\frac{1}{2} R_{\rm in} ^2 } {\frac{1}{R_{\rm in}}-\frac{2 R_{\rm c}^{1/2}}{3 R_{\rm out}(t)^{3/2}}-\frac{1}{3 R_{\rm c}}}
   \label{eq:Tprec}
\end{equation}
Since only the geometrically-thick portion of the disk will precess as a solid-body, we estimate\footnote{The trapping radius $R_{\rm Edd}$ may slightly underestimate the radius where solid body precession ceases ($H/r \lesssim \alpha$), particularly if $\alpha$ is small; however, the outermost regions of the disk where $H/r \sim \alpha$ will expand viscously slower than estimated by Eq.~\eqref{eq:Rd} assuming $H/r \sim 0.3$, in part compensating for this error.  Additional uncertainty enters because radiation-dominated $\alpha$-disks without advective cooling are known to be thermally- and viscously-unstable.}
\be
R_{\rm out} = {\rm min}[R_{\rm d},R_{\rm Edd}]
\label{eq:Rout2}
\ee
to be the minimum of the outer edge of the spreading disk (Eq.~\ref{eq:Rd}) and the sub-Eddington radius (Eq.~\ref{eq:Redd}).

\begin{figure*}
    \centering
    \includegraphics[width=0.45\textwidth]{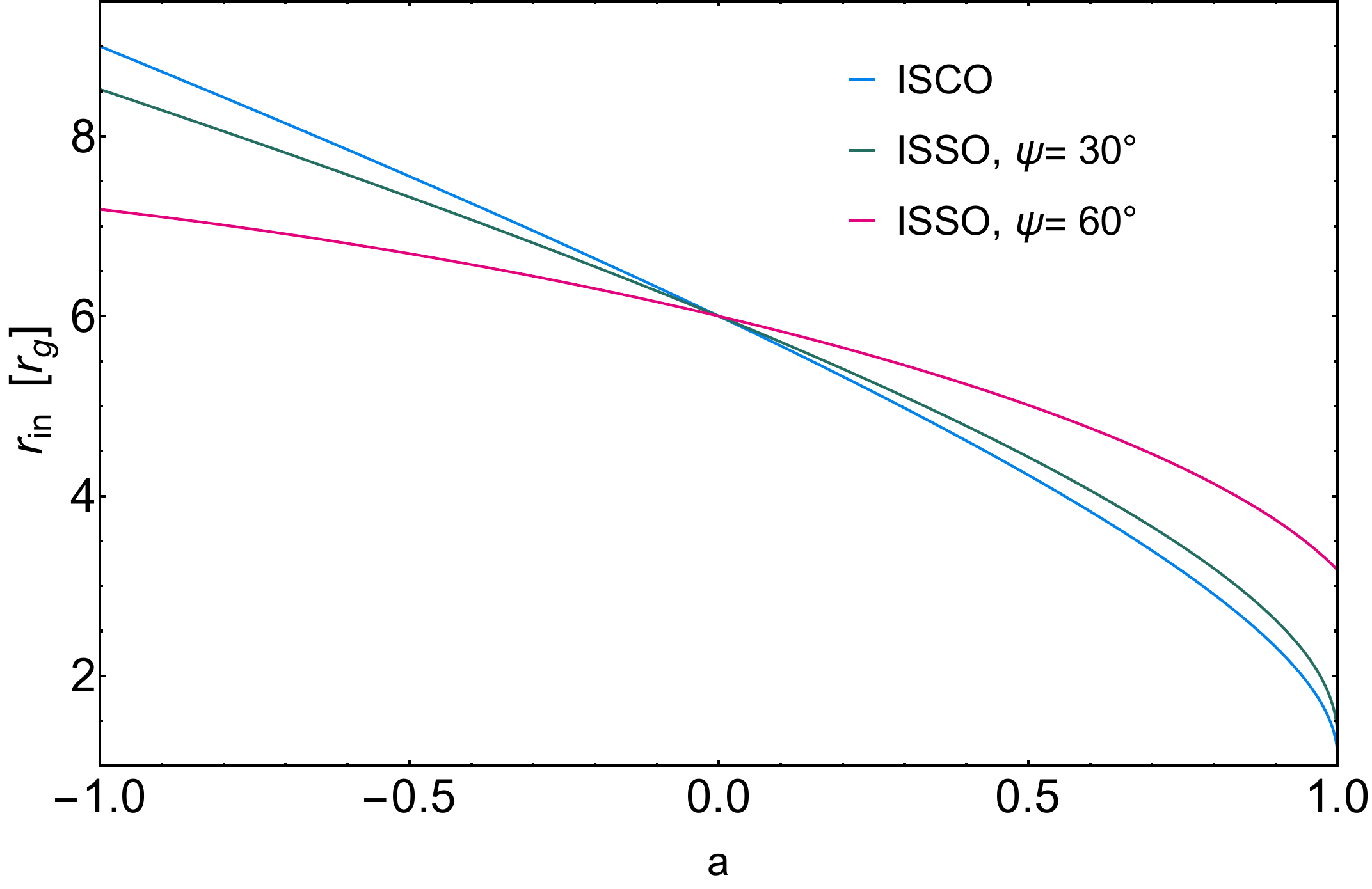}
    \includegraphics[width=0.48\textwidth]{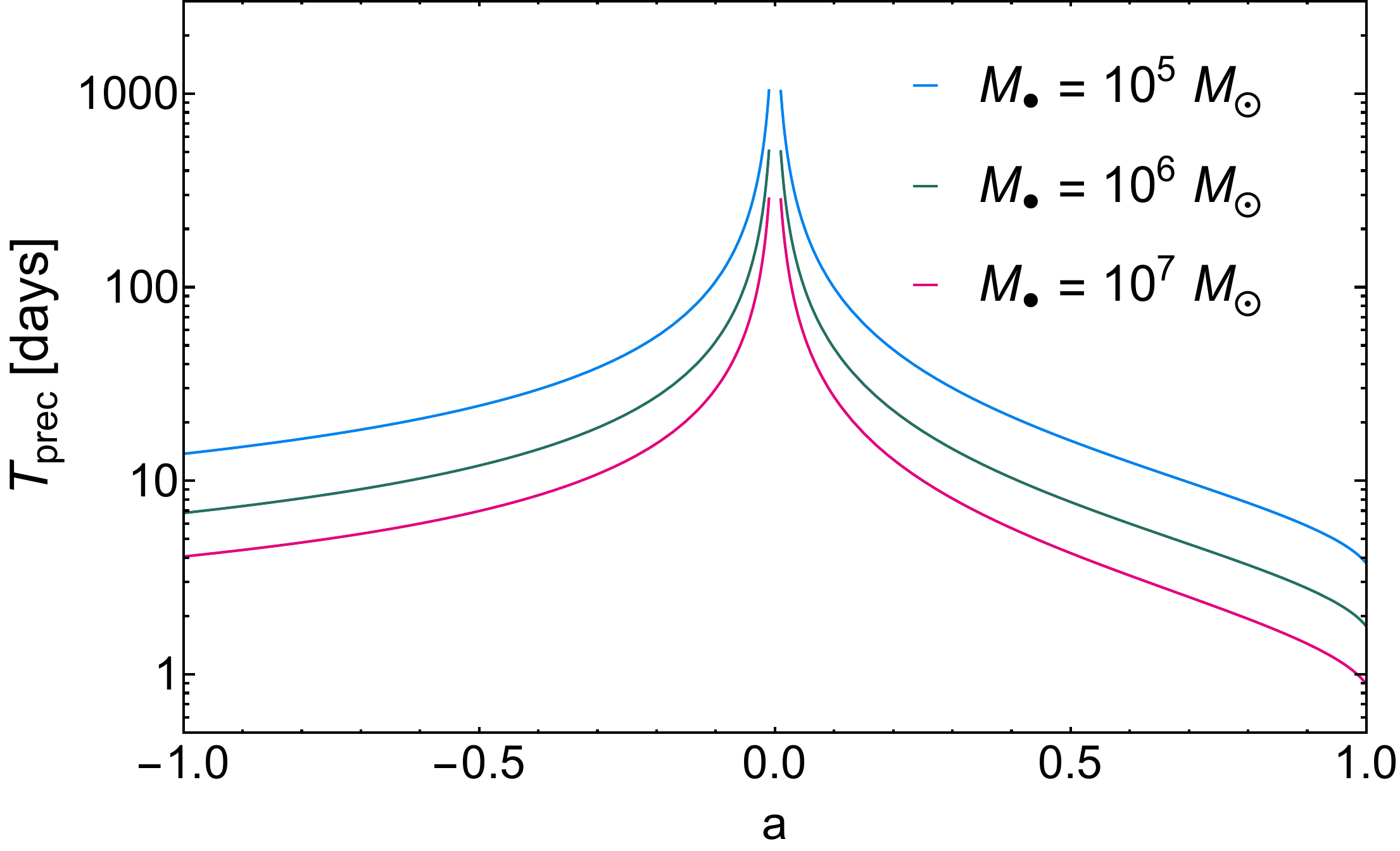}
    \includegraphics[width=0.48\textwidth]{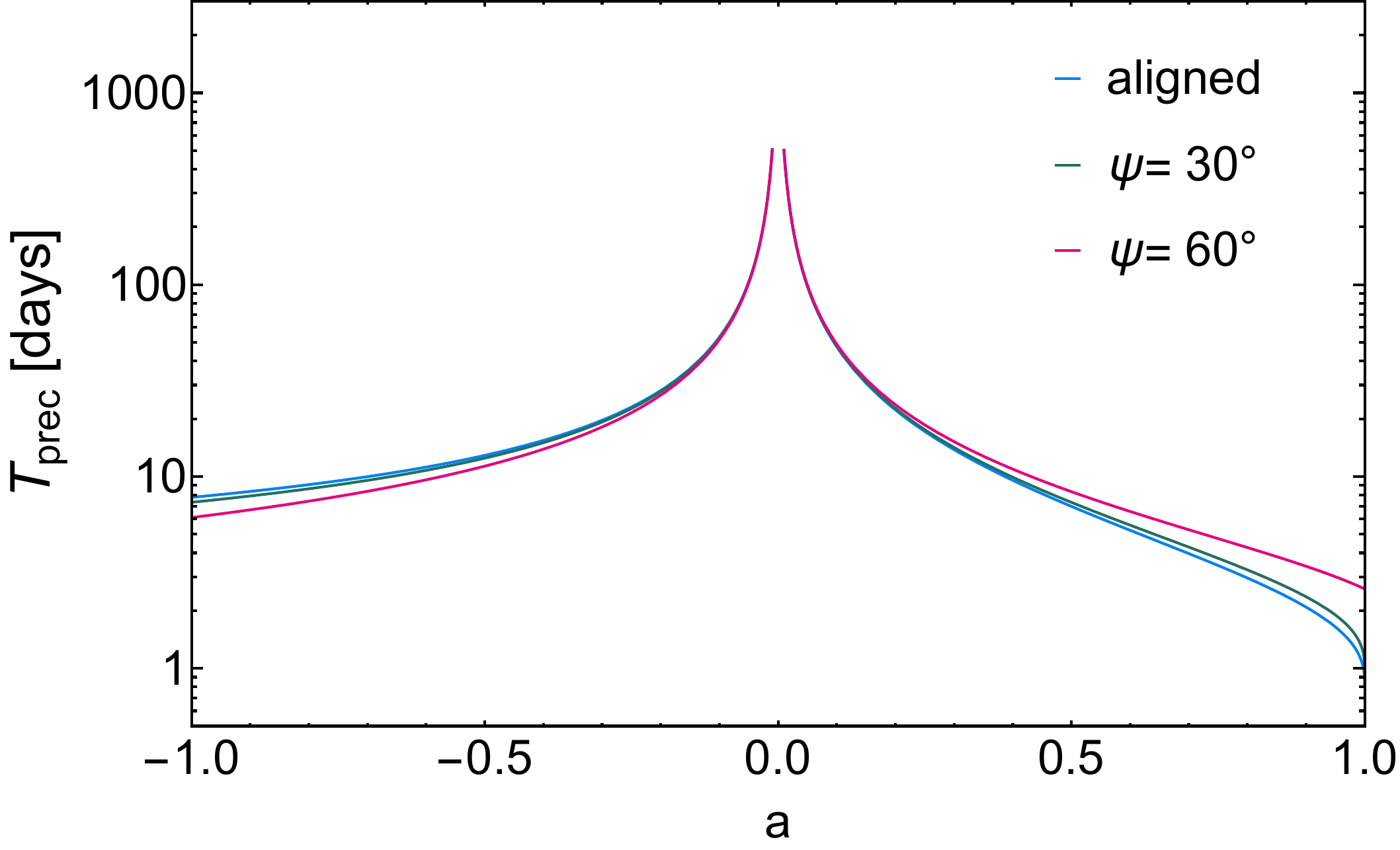}
    \includegraphics[width=0.48\textwidth]{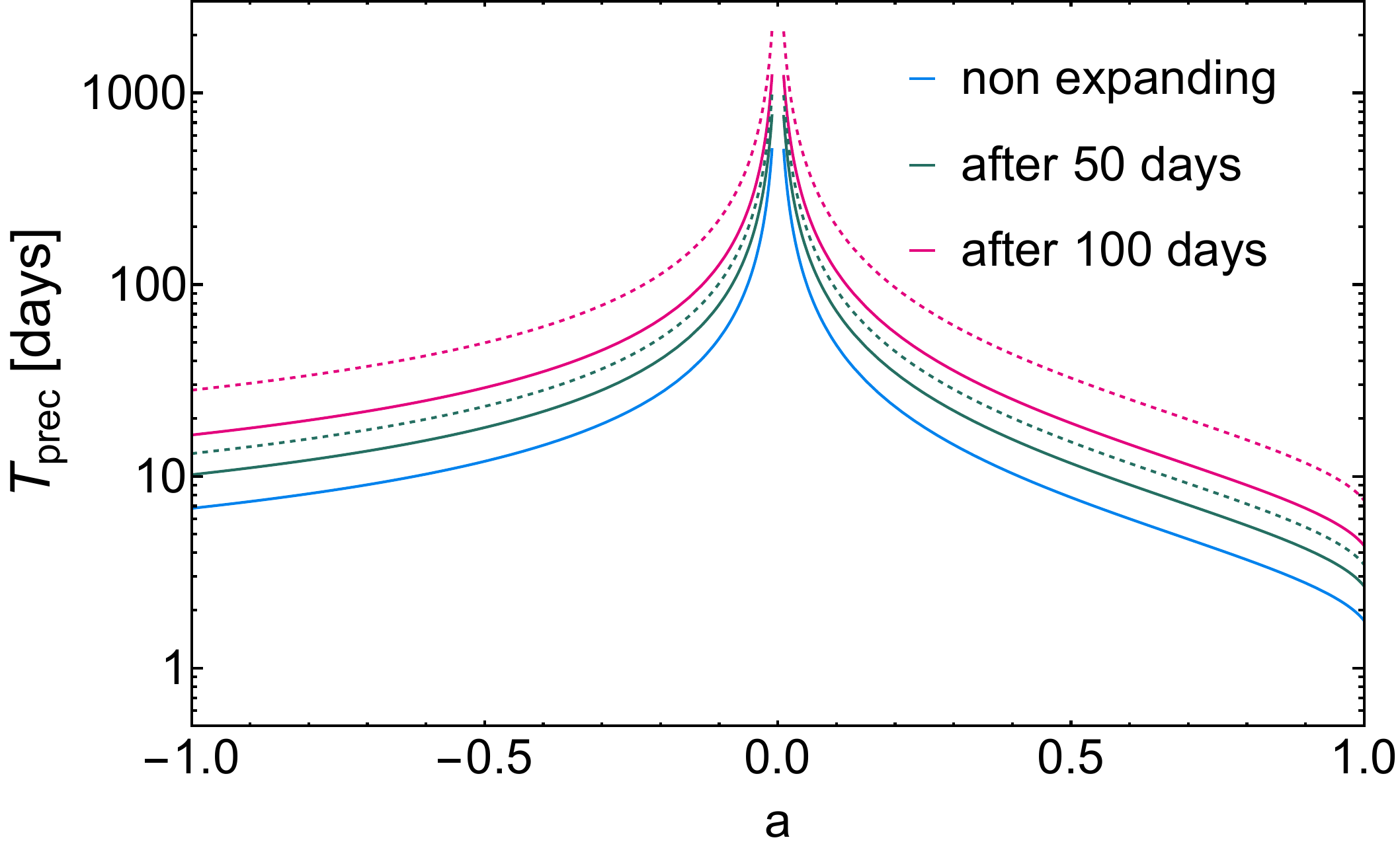}
    \caption{{\it Top left panel:} Inner radius of the accretion disk at either the ISCO or ISSO in units of gravitational radii, as a function of the SMBH spin $a$, for different values of the spin-disk misalignment angle $\psi$ as marked. {\it Other panels:} Precession period of the disk created by the tidal disruption of a Sun-like star as a function of the SMBH spin $a$, with different panels showing the separate effect of changing: (a) the SMBH mass $M_{\bullet}$ (for fixed $\psi = 30^{\circ}$, $R_{\rm out} = R_{\rm c}$); (b) misalignment angle $\psi$ (for fixed $M_{\bullet} = 10^{6}M_{\odot}$ and $R_{\rm out} = R_{\rm c}$); (c) the outer radius $R_{\rm out}$ (for fixed $M_{\bullet} = 10^{6}M_{\odot}$, $\psi = 30^{\circ}$), particularly whether $R_{\rm out} = R_{\rm c}$ is fixed or whether the disk is allowed to expand viscously.  In the latter case, solid lines correspond to the fiducial spreading evolution taking into account wind angular momentum losses $R_{\rm d} \propto t^{2/5}$ (Eq.~\eqref{eq:Rd}) while dashed lines correspond to the fiducial expansion rate without wind angular momentum losses $R_{\rm out} \propto t^{2/3}$.}
\label{fig:Tprec}
\end{figure*}

The three panels of Fig.~\ref{fig:Tprec} show the disk precession time $T_{\rm prec}$ for the disruption of a solar mass star as a function of SMBH spin $a$, for different assumptions about the SMBH mass, disk-spin misalignment angle $\psi$, and the value of $R_{\rm out}$ (in particular, whether we fix $R_{\rm out} = R_{\rm c}$ or allow the disk to spread according to Eq.~\eqref{eq:Rd}, in which case $T_{\rm prec}$ becomes time-dependent).  For moderately high SMBH spin $|a| \gtrsim 0.5$ the precession period is typically of order days, though it becomes longer for less-massive SMBH or in the case of retrograde spin.  By contrast, for a fixed SMBH mass, the precession period depends only weakly on the misalignment angle because $\psi$ only enters through $R_{\rm in}$, while $T_{\rm prec}$ is more sensitive to $R_{\rm out}$.  Indeed, in cases where $R_{\rm out} \approx R_{\rm d}$ is assumed to expand at the maximal rate $\propto t^{2/3}$ (Eq.~\eqref{eq:Rd}), $T_{\rm prec}$ grows significantly over the first hundred days after the disruption.  On the other hand, if the outer radius grows more gradually with time as expected for wind angular-momentum losses, the change in $T_{\rm prec}$ is more modest.  

In summary, for the moderate to high-spin cases $|a| \gtrsim 0.5$ most likely to launch powerful jets via the BZ mechanism, the precession timescales are typically days to weeks, and much smaller than the sub-Eddington transition $t_{\rm Edd}$ (Eq.~\eqref{eq:tEdd}). 

\subsection{Disk/Jet Alignment}
\label{sec:alignment}

Although the angular momentum axis of the accretion disk, and hence of the black hole jet, will in general be initially misaligned with the SMBH spin, different mechanisms can act to align the two over time following the TDE (e.g., \citealt{Papaloizou&Pringle83,McKinney+13,Franchini+16}).  We first consider ``hydrodynamic'' alignment mechanisms, which act as a result of internal stresses within the accretion disk itself, before considering alignment due to large-scale magnetic stresses on the disk by the jet.  

\subsubsection{Hydrodynamic Alignment}

At early times when $\dot{M} \gg \dot{M}_{\rm Edd}$, radiation is trapped and radially advected through the disk (e.g., \citealt{Begelman79}), rendering it hot and geometrically thick, with $H/r \sim 0.3 \gtrsim \alpha$ (e.g., \citealt{Sadowski&Narayan15}).  Such geometrically thick disks with $H/r \gtrsim \alpha$ precess as solid bodies about the spin axis of the SMBH (e.g., \citealt{Fragile+07}).  However, as $\dot{M} \approx \dot{M}_{\rm fb}$ drops, the disk will eventually transition to a thinner, radiatively-cooled state with $H/r \lesssim \alpha$, enabling it to undergo alignment with the SMBH spin through the \citet{Bardeen&Petterson75}, eventually out to radii $R_{\rm BP} \simeq 30\alpha_{-1}^{-4/7}R_{\rm in} \gtrsim R_{\rm c}$ encompassing most of the disk (e.g., \citealt{Nelson&Papaloizou00}). 

For disks with fixed aspect ratio $H/r$, the Bardeen-Petterson effect which causes disk alignment with the BH spin axis usually first occurs at the small disk radii closest to the black hole and moves outwards with time.  However for the evolution of the disk surface density we have derived (Eq.~\ref{eq:Sigma_analytic}), the outer portions of the TDE disk become geometrically-thin first ($r \gtrsim R_{\rm Edd}$).  We hypothesize this causes the alignment to begin there first and using Eq.~\eqref{eq:Rd} for the outer edge of the disk $R_{\rm d}(t)$, we estimate that the outermost disk radii will stop precessing as a solid body (i.e., $(H/r)|_{R_{\rm d}} = \alpha$), after a time
\begin{eqnarray}
t_{\rm align}^{\rm SB} \simeq \left[\frac{10R_{\rm g}}{\alpha R_{\rm c}}t_{\nu,\rm c}^{\chi}t_{\rm Edd}^{5/3}\right]^{\frac{3}{5+3\chi}} \underset{\chi = 2/3}\simeq 193\alpha_{-1}^{-23/21}M_{\bullet,6}^{4/35}\,{\rm d}, \nonumber \\
\label{eq:talign1}
\end{eqnarray}
where in the final line we take $\chi = 2/3$ and have used Eqs.~\eqref{eq:Rt}, \eqref{eq:tnuc}, \eqref{eq:tEdd}, as well as the fact that the scale height of a radiation-dominated disk is given by $H \simeq 10R_{\rm g}(\dot{M}/\dot{M}_{\rm Edd})$ (e.g., \citealt{Frank+02}).  The outermost radius of the disk is crucial for the jet evolution as it determines the jet direction on large scale (e.g., \citealt{Liska+18}) insofar that disk winds collimate the jet.   

Thus, if other mechanisms do not operate sooner, jet alignment could occur as quickly as several months, depending on the radial expansion rate of the disk and the sensitivity of the large-scale jet-alignment process to the radial profile of $H/r.$  If alignment with the black hole spin axis is not immediate, the lack of out disk precession would at least quench the jet's precession, effectively fixing its drilling angle on timescales relevant to the jet propagation.
Such a configuration of an aligned outer axis with inner geometrically-thick  precessing about the spin axis has been found in GRMHD simulation  albeit in a different physical context (e.g., see \citealt{Bollimpalli+23}), though this merits additional numerical studies.

As first pointed out by \citet{Franchini+16}, another hydrodynamic alignment can occur when $a$ and $\alpha$ are both sufficiently large.  In this regime, internal stresses within a thick, rigidly precessing disk will act to slowly damp out the misalignment angle \citep{Bate+00,Foucart&Lai14}.  For different assumptions about the disk viscosity $\alpha \sim 10^{-2}-10^{-1}$ and SMBH spin, \citet{Franchini+16} estimate alignment timescales through this mechanism for a 10$^{6}M_{\odot}$ SMBH ranging from several several hundred to several thousand days (see their Fig.~9).  Although the physical mechanism is different, these values are comparable to our estimate above (Eq.~\eqref{eq:talign1}) for the alignment time arising due to the thinning of the outer spreading disk.

\subsubsection{Magneto-Spin Alignment}
\label{sec:magnetospin}

For sufficiently powerful magnetized jets, another alignment process can act faster than purely hydrodynamical processes.  In this ``magneto-spin alignment'' process, torques from the electromagnetic (EM) fields threading the jet act on the disk and align it with the SMBH spin axis \citep{McKinney+13,Polko+17}. Following \citet{McKinney+13}, the radius $R_{\rm ms}$ at which the magneto-alignment occur can be estimated by equating the EM torque with the torque per unit area of the accreting plasma : $ r B_r B_{\phi}/4 \sim \dot{M}\Omega/2\pi$. Here, $B_{\phi} \sim r B_r \Omega_{\rm H}$ is the toroidal magnetic field of the jet and $\Omega_{\rm H}$ is the angular frequency of the SMBH and hence 
\be
R_{\rm ms} \sim \frac{c}{v_{\phi}}\omega_{\rm H}\Upsilon^{2}R_{\rm g},
\ee
where $v_{\phi} = r\Omega$ and again $\omega_{\rm H}= a/(1+(1-a^{2})^{1/2}) $ is the dimensionless SMBH angular frequency and $\Upsilon$ its magnetic flux.  Alignment ($R_{\rm ms} > R_{\rm c}$) with the SMBH spin is therefore expected above a critical magnetic flux,
\be
\Upsilon \gtrsim \Upsilon_{\rm align} \approx \left(\frac{4 R_{\rm t}^{2}\Omega}{\beta^{2} c \omega_{\rm H}R_{\rm g}}\right)^{1/2} = \frac{2}{\beta\omega_{\rm H}^{1/2}}\left(\frac{R_{\rm t}}{ R_{\rm g}}\right)^{1/4},
\label{eq:Upsilonalign}
\ee
or, equivalently, above a BZ jet efficiency $\eta_{\rm BZ}$ (Eq.~\eqref{eq:etaBZ}) 
\be
\eta_{\rm BZ,align}  \approx \frac{0.08}{\beta^{2}} \left(\frac{R_{\rm t}}{R_{\rm g}}\right)^{1/2}\omega_{\rm H}f(\omega_{\rm H}) \approx \frac{0.54}{\beta^{2}} M_{\bullet,6}^{-1/3}\omega_{\rm H}f(\omega_{\rm H}).
\label{eq:etaBZalign}
\ee
The requirement that the magnetic flux for alignment $\Upsilon_{\rm align}$ not exceed the maximum allowed value $\Upsilon_{\rm max} \sim 20$ ($\eta_{\rm BZ} \sim 1$) corresponding to MAD accretion, places a lower limit on the black hole spin for alignment,
\be
\omega_{\rm BH,min} \sim a_{\rm min} \sim 0.07\beta^{-2}(R_{\rm t}/R_{\rm g})^{1/2} \sim 0.07\beta^{-2}M_{\bullet,6}^{-1/3}.
\label{eq:omegaBHmin}
\ee

Once the SMBH has accreted sufficient magnetic flux for $\eta \gtrsim \eta_{\rm BZ}$, the actual alignment process will be quick, occuring on the timescale,
\be
t_{\rm j} \sim \frac{\Sigma_{\rm c}j_{\rm c}}{\tau_{\rm acc}} \sim \frac{R_{\rm c}^{2}}{\nu_{\rm c}} \sim t_{\nu, \rm c},
\ee 
where $j_{\rm c} = R_{\rm c}^{2}\Omega$ is the specific angular momentum at the disk radius $\sim R_{\rm c}$.  Thus, the jet magneto-spin alignment time is given by
\be
t_{\rm align,mag} = t_{\Phi} + t_{\nu, \rm c} \simeq t_{\Phi},
\label{eq:talignmag}
\ee
where $t_{\Phi}$ is the time to accumulate the critical amount of magnetic flux $\Upsilon_{\rm align}$ 
(Eq.~\eqref{eq:Upsilonalign}).

\subsubsection{Comparison of Jet Alignment Mechanisms}

\begin{figure*}
\centering
\includegraphics[width=0.7\textwidth]{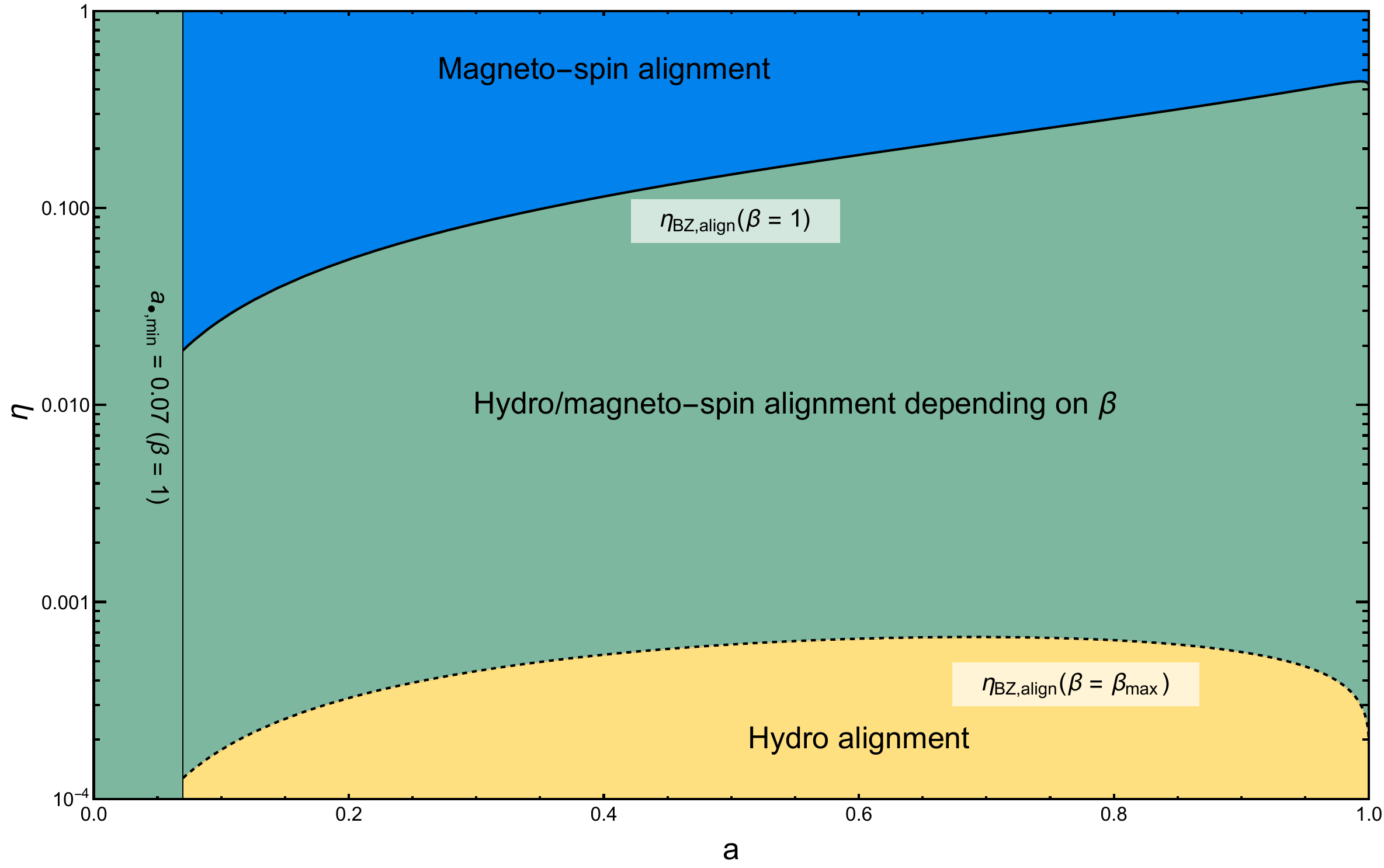}
\caption{Regimes of distinct jet alignment mechanisms (``magneto-spin'' versus ``hydrodynamical'') in the parameter space of dimensionless SMBH spin $a$ and jet efficiency $\eta$ for $M_\bullet = 10^6 M_\odot$, disruption of a solar-type star and a typical disk-spin misalignment angle $\psi =30^{\circ}$.  For sufficiently powerful jets powered by the BZ process ($\eta \gtrsim \eta_{\rm BZ,align}$; Eq.~\eqref{eq:etaBZalign}), the magnetic field is sufficiently strong that magneto-spin alignment is the fastest process, occurring on a timescale as short as the fall-back time over which most of the mass and magnetic flux is assumed to be accreted; we show this boundary separately for $\beta = 1$ and $\beta = \beta_{\rm max}$ (Eq.~\eqref{eq:betamax}).  Our estimate of the minimum SMBH spin required for magneto-spin alignment (Eq.~\eqref{eq:omegaBHmin} for $\beta = 1$) is also shown with a vertical line.}
\label{fig:alignment}
\end{figure*}

Depending on the jet efficiency $\eta$, the SMBH spin $a$ and the impact parameter $\beta$ of the disrupted star, the disk/jet system will align first through either the magneto-spin or the hydrodynamic mechanisms, as summarized in Fig.~\ref{fig:alignment}.  We find that the most powerful jets with $\eta \gtrsim 10^{-2}-10^{-1}$ will undergo magneto-spin alignment, while weaker jets ($\eta \lesssim 10^{-3}$) will align through an hydrodynamic mechanism.  For intermediate jet efficiency, either alignment mechanism can operate, depending on $\beta$ and the spin parameter $a$. 

When magneto-spin alignment occurs, it generally takes place much faster than hydrodynamic alignment.  The magneto-spin alignment timescale $t_{\rm align,mag} \approx t_{\Phi}$ (Eq.~\eqref{eq:talignmag}) depends on the origin of the jet's magnetic flux, which is not well constrained. If the magnetic flux originates from inside the disrupted star (e.g., \citealt{Bonnerot+17,Bradnick+17}), then most of the total magnetic flux is accumulated when most of the star's mass is accreted on the fall-back time, i.e.~$t_{\Phi} \sim t_{\rm fb}.$  On the other hand, if the magnetic flux is dragged into the SMBH from the pre-TDE circumnuclear environment (e.g., from a ``fossil'' AGN accretion disk; \citealt{Kelley+14}) by the infalling stellar debris streams, then \citet{Tchekhovskoy+14} predict $\Phi_{\bullet} \propto t^{2/3}$ will grow with time as the apocenter radius of the debris stream expands to encompass more of the fossil disk and hence the dimensionless flux $\Upsilon \propto \Phi_{\bullet}/\sqrt{\dot{M}_{\rm fb}} \propto t^{3/2}$ will grow rapidly to late times $t \gg t_{\rm fb}$.  In this case, the time for the magnetic flux to saturate $t_{\Phi}$, or for the SMBH to approach the MAD state, could in principle be $\gg t_{\rm fb},$ depending on the properties of the fossil disk.  

In summary, powerful and mildly powerful jets with high $\beta$ will undergo magneto-spin alignment on a timescale of $\lesssim t_{\rm fb} \sim$ weeks or potentially somewhat longer, depending on the magnetic field of the disrupted star, the efficiency of a magnetic dynamo in the accretion disk, and (potentially) the pre-existing circumnuclear environment.  By contrast, weak and mildly powerful jets with $\beta \sim 1$ will undergo delayed alignment through the hydrodynamic mechanism on a much longer timescale of many months to years (Eq.~\eqref{eq:talign1}; Fig.~\ref{fig:alignment}).

\section{Jet Propagation}
\label{sec:propagation}
Depending primarily on the properties of the SMBH, such as its spin and magnetic flux, a bipolar relativistic jet can be launched along the angular momentum axis of the accretion disk (Sec.~\ref{sec:jet}). Our results in the previous section show that the jet is generally expected to undergo rapid precession around the SMBH spin axis (Sec.~\ref{sec:precession}) on a timescale typically much shorter than that required for the jet and spin axis to become aligned (Sec.~\ref{sec:alignment}). This rapid precession has two effects: i) the precessing jet shares its energy with a larger conical with half-opening angle $\psi$, the misalignment angle, ii) the jet head follows an helical motion (Fig.~\ref{fig:cartoon}). 

 Here we consider the propagation of such a relativistic jet in the dense wind material released from the super-Eddington disk. In Sec.~\ref{sec:wind}, we estimate the properties of the precession-driven wind.  In Sec.~\ref{sec:jet head} we calculate the conditions for the jet to successfully propagate through the wind for both a precessing and a non-precessing jet. 

\subsection{TDE Envelope from Precession-Driven Wind}
\label{sec:wind} 

The super-Eddington accretion phase following a TDE is expected to be accompanied by mass-loaded disk winds with a wide range of outflow speeds $v \sim 0.01-0.3$ c (e.g., \citealt{Strubbe&Quataert09,Metzger&Stone16,Dai+18,Dai+21}).  In support of this picture, synchrotron emission from non-relativistic outflows of velocity $v \gtrsim 10^{4}$ km s$^{-1}$, consistent with being launched around the time of disruption, have been detected from a handful of nearby TDEs (e.g., \citealt{Alexander+17,Alexander+20}).  Insofar that the majority of the detected sample represent the closest events for which this relatively weak radio flux (compared to the radio afterglows of a relativistic TDE jets) is detectable, it is likely that most if not all TDEs are accompanied by such outflows.  

We assume that a fraction $f_{\rm w} < 1$ of the mass fall-back rate to the SMBH, $\dot{M}_{\rm fb}$ is launched from the disk across a range of radii $R_{\rm w} \lesssim R_{\rm out}$ with a mass-averaged speed $v_{\rm w} = \beta_{\rm w}c$.  As a result of the rapid precession of the disk angular momentum axis (Sec.~\ref{sec:precession}), these disk winds will inevitably interact and collide with each other on large scales $\gg R_{\rm w}$ (see Fig.~\ref{fig:cartoon}).  This will encase the inner disk in a quasi-spherical outflow, with a radial density profile which on large scales can be approximated as:
\begin{equation}
\rho_{\rm w} =\frac{f_{\rm w} \dot{M}_{\rm fb}}{4 \pi v_{\rm w} r^2}, \quad R_{\rm min} <r<R_{\text {\rm max}}.
\label{eq:rhow}
\end{equation}
A rough upper limit on the innermost radius of the quasi-steady wind zone is given by the distance the wind traverses in a single precession period:
\be
R_{\rm min} \leq v_{\rm w}T_{\rm prec} \sim  10^{15}\beta_{\rm w,-1}\left(\frac{T_{\rm prec}}{3\,{\rm d}}\right)\,{\rm cm}\,,
\label{eq:Rmin}
\ee
where $\beta_{\rm w,-1} = \beta_{\rm w}/(0.1)$ and we have normalized $T_{\rm prec}$ to a typical value (Fig.~\ref{fig:Tprec}).  The maximum radius of the steady wind zone at a given time $\sim t \gtrsim t_{\rm fb}$ can likewise be estimated as the disk the wind traverses in time $\sim t$ since disruption:
\begin{equation}
R_{\rm max} \sim v_{\mathrm{w}} t \sim 1.5 \times 10^{16} M_{\bullet,6}^{1/2}\beta_{\rm w,-1}\left(\frac{t}{t_{\mathrm{fb}}}\right) \,{\rm cm}\,.
\label{eq:Rmax}
\end{equation}

The Thomson optical depth through the wind external to a radius $r>R_{\text {in }}$ is given by
\begin{eqnarray}
\tau &\simeq& \frac{f_{\rm w}\dot{M}\kappa_{\rm es}}{4 \pi v_{\rm w} r} \nonumber \\
&\simeq& 0.2 M_{\bullet,6}^{-1/2}\frac{f_{\rm w,-1}}{\beta_{\rm w,-1}^{2}}\left(\frac{T_{\rm prec}}{3\,{\rm d}}\right)^{-1}\left(\frac{r}{R_{\rm min}}\right)^{-1}\left(\frac{t}{t_{\rm fb}}\right)^{-5/3},
\end{eqnarray}
where $\kappa_{\rm es} \simeq 0.34$ cm$^{2}$ g$^{-1}$ and we have used Eq.~\eqref{eq:Rmin}. The photon diffusion timescale outwards through the wind from radius $r$ is given by $t_{\rm diff} \sim \tau(r/c)$, such that photons are ``trapped" $\left(t_{\rm diff}>t_{\rm exp} \sim r/v_{\rm w}\right)$ at radii interior to
\begin{eqnarray}
R_{\rm trap} &\simeq& \frac{f_{\rm w}\dot{M} \kappa_{\rm es}}{4 \pi c} \nonumber \\
&\approx& 1.4 \times 10^{13}f_{\rm w,-1}\left(\frac{t}{t_{\rm fb}}\right)^{-5/3}\,{\rm cm}\, \ll R_{\rm max}
\label{eq:Rtrap}
\end{eqnarray}
The fact that $R_{\rm trap} > R_{\rm min}$ shows that the shock interaction between the jet and the wind-envelope (which takes place at radii $> R_{\rm min}$) will be mediated by collisionless plasma processes, rather than by trapped radiation.  It also shows that any radiation from a trapped jet (e.g., reprocessed UV/optical emission) will escape from the wind ejecta without experiencing significant adiabatic expansion losses.    

\subsection{Jet Head Motion}
\label{sec:jet head}

We consider a relativistic jet with a kinetic luminosity $L_{\rm j}$ (Eq.~\eqref{eq:Ljet}), bulk Lorentz factor $\Gamma_{\rm j} \gg 1$ and half-opening angle $\theta_{\rm j}$, such that its isotropic kinetic luminosity within the jet opening cone follows $L_{\rm iso}= 2 L_{\rm j}/ \theta_{\rm j}^2$. 

The expansion velocity of the jet head is given to good approximation by the balance of jet and ambient ram pressures in the frame of the jet head:
\begin{equation}
\rho_{\rm j} h_{\rm j} ( \Gamma \beta)^2_{\rm jh} c^2 + p_{\rm j} = \rho_{\rm w} h_{\rm w} ( \Gamma \beta)^2_{\rm h} c^2+ p_{\rm w}, 
\label{eq:pressurebalance}
\end{equation}
where $(\Gamma \beta)_{\rm jh} $ is the relative four-velocity between the jet and its head $(\Gamma \beta)_{\rm h}$, while $\rho, p$ and $h$ are the density, pressure and enthalpy, respectively, where as usual the subscripts ``j'' and ``w'' refer to the jet and wind, respectively.
\\
Approximating the slowly expanding upstream wind as a stationary medium, we have:
\begin{equation}
\beta_{\rm h}= \frac{\beta_{\rm j}}{1+  \tilde{L} ^{-1/2}},
\label{eq:betah}
\end{equation}
where the dimensionless jet luminosity \citep{Marti+94,Matzner03} is:
\begin{equation}
 \tilde{L} \equiv \frac{\rho_{\rm j} h_{\rm j}  \Gamma_{\rm j}^2 }{\rho_{\rm w}} = \frac{L_{\rm iso}/c^3}{dM_{\rm w}/dr} = \frac{2\eta}{\theta_{\rm j}^{2}\beta_{\rm w}},
 \label{eq:tildeL}
\end{equation} 
where $M_{\rm w}$ is the mass of the wind.  Following the discussion in Sec.~\ref{sec:wind}, we consider the environment surrounding the SMBH at large radii $r \geq R_{\rm min}$ as a spherical wind of constant velocity $\beta_{\rm w} c$ and density profile $\rho_{\rm w} \propto r^{-2}$ (Eq.~\eqref{eq:rhow}) leading to $dM_{\rm w}/dr = f_{\rm w}\dot{M}/v_{\rm w}$, where the final line of Eq.~\eqref{eq:tildeL} also makes use of $L_{\rm iso} = 2L_{\rm j}/\theta_{\rm j}^{2}$

In either of two limits, $\tilde{L} \ll 1$ or $1 \ll \tilde{L} \ll \Gamma_{\rm j}^4$, Eq.~\eqref{eq:betah} can be simplified to give the velocity of the jet head:
 \begin{equation} 
\beta_{\rm h} \approx \tilde{L}^{1/2}.
\label{eq:escape}
\end{equation}

We consider the jet propagation under two scenarios. In the case of a non-precessing jet initially launched in the SMBH spin axis, or precessing but aligned to within an angle $\lesssim \theta_{\rm j}$ of the SMBH spin axis the condition \eqref{eq:escape} can be written
\begin{equation}
\tilde{L} \gtrsim \beta_{\rm w}^{2}.
\label{eq:success}
\end{equation}

Using definition \eqref{eq:tildeL}, Eq.~\eqref{eq:success} can be expressed as a lower limit on the jet efficiency: 
\begin{eqnarray}
&& \eta > \eta_{\rm crit,1} \approx \frac{f_{\rm w}\theta_{\rm j}^{2}\beta_{\rm w}}{2} \nonumber \\
&\simeq& 5\times 10^{-5}f_{\rm w,-1}\beta_{\rm w,-1}\left(\frac{\theta_{\rm j}}{0.1}\right)^{2}\,\text{(Non-Precessing)}\,\,\,\,
\label{eq:etacritslow}
\end{eqnarray}
For jets either initially aligned with SMBH spin axis, or those which have underwent alignment after some delay, even small efficiencies are sufficient to escape from the TDE wind environment.  

However, as emphasized earlier, only a very small fraction of TDE jets begin aligned at the time of disruption. Furthermore, in the more common misaligned cases, the timescale $\sim \theta_{\rm j}T_{\rm prec} \sim 0.1T_{\rm prec}$, over which the jet will precess across an angle larger than the jet opening angle $\sim \theta_{\rm j}$ (Sec.~\ref{sec:precession}), is typically short compared to that required for the jet to escape from the wind $\gtrsim R_{\rm min}/(\beta_{\rm j}c)$.  

The vast majority of TDE jets are misaligned and undergo rapid precession on a timescale much faster than the alignment time (Sec.~\ref{sec:alignment}), leading to a more stringent criterion for jet escape.  As illustrated by the inset within Fig.~\ref{fig:cartoon}, the jet head follows an helical movement with velocity $\vec{v} = (\mathcal{\rho} \dot{\phi})\hat{\phi} + (H\dot{\phi}/2\pi n)\hat{z}$ in cylindrical coordinates ($\mathcal{\rho},\phi,z)$, where $\dot{\phi}$ is the azimuthal speed, $H$ is the pitch of the helix and $n$ the number of precession cycles.  Defining the misalignment angle according to $\tan \psi = \mathcal{\rho}/(nH)$, the total magnitude of the head velocity can be written 
\be |v_{\rm h}|= \frac{H \dot{\phi}}{2 \pi}\sqrt{4 \pi^2 n^2 \tan^2(\psi)+1},
\ee
The propagation velocity of the head along the fixed SMBH spin axis is then given by:
\begin{equation}
v_{\rm h,z} = \frac{|v_{\rm h}|}{\sqrt{1 + 4 \pi^2 \tan^2 \psi}},
\end{equation} 
and the condition for successful jet escape in the rapid precession case ($\beta_{\rm h,z} \gtrsim \beta_{\rm w}$) can be written: 
\begin{equation}
 \tilde{L} \geq \beta_{\rm w}^{2}(1 + 4 \pi^2 \tan^2 \psi), \,\,\, \text{(Precessing).} 
 \label{eq:success2}
\end{equation}   
A rapidly precessing jet effectively shares its energy with a much larger conical structure: a jet of half opening angle $\psi$ roughly equal to the misalignment angle (see Fig.~\ref{fig:cartoon}); hence, the minimum efficiency for a successful jet in the precessing/misaligned case becomes
\begin{equation}
    \eta_{\rm crit,2} \approx \frac{\psi^2 f_{\rm w} \beta_{\rm w}}{2}\left(1 + 4 \pi^2 \tan^2 \psi \right) \,\,\,\text{(Precessing)}.\label{eq:etacritfast}
\end{equation}
We note that $\eta_{\rm crit,2}$ is typically much higher (by a factor $\gtrsim \psi^{2}/\theta_{\rm j}^{2} \left(1 + 4 \pi^2 \tan^2 \psi \right) $) than in the non-precessing case ($\eta_{\rm crit,1}$; Eq.~\eqref{eq:etacritslow}).

\begin{figure}
\centering
\includegraphics[width=0.45\textwidth]{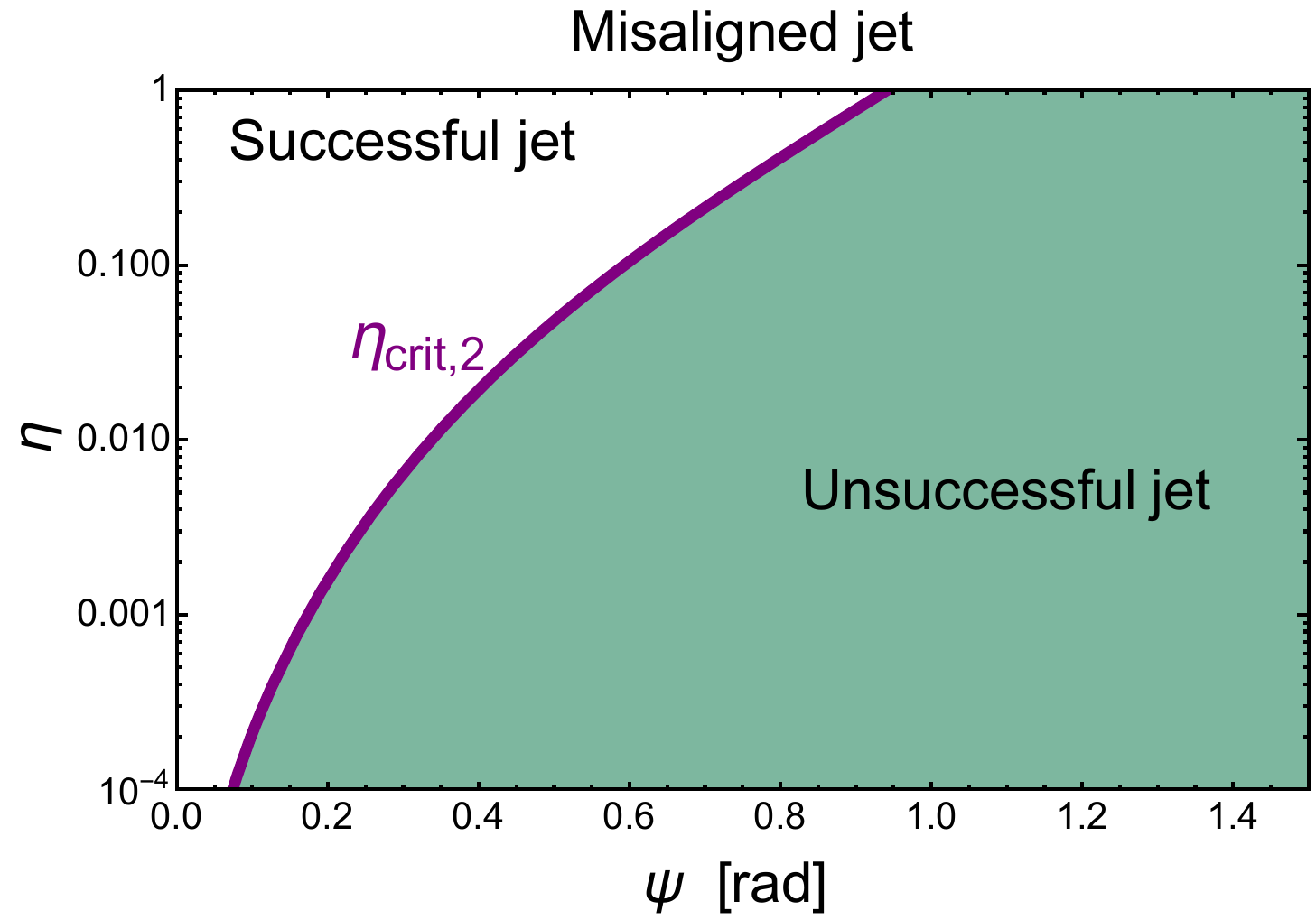}

\includegraphics[width=0.47\textwidth]{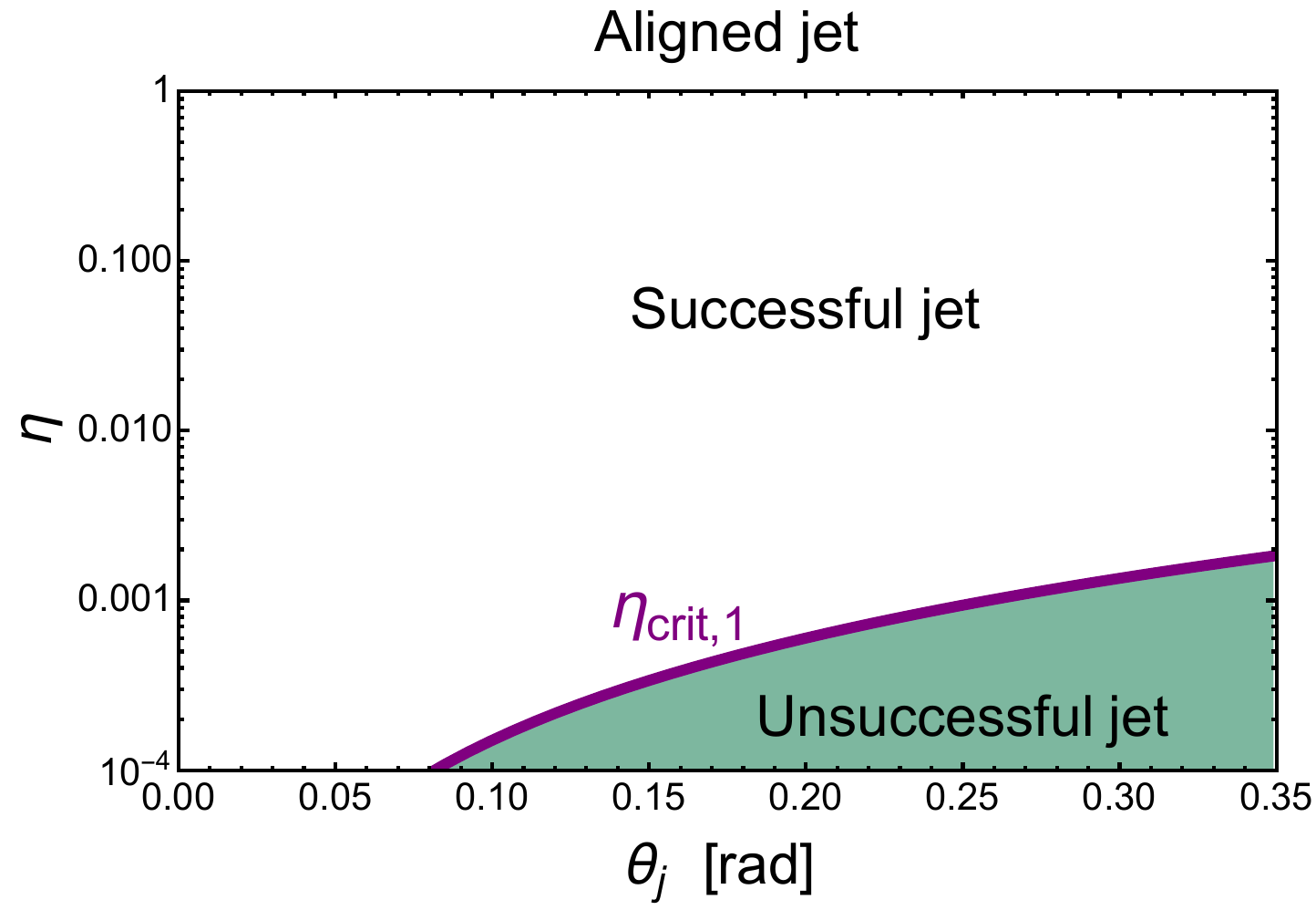}
\caption{Parameter space of jet efficiency $\eta$ versus disk/jet-spin misalignment angle $\psi$ or jet half-opening angle $\theta_{\rm j}$ required for successful escape of the TDE jet from the disk-wind envelope for $f_{\rm w} =0.15$, $\beta_{\rm w}=0.2$.  We show separately cases for a rapidly precessing misaligned jet (top panel) and an aligned jet (bottom panel).  The critical minimum efficiency for a successful jet before ($\eta_{\rm crit,1}$; Eq.~\ref{eq:etacritfast}) and after alignment ($\eta_{\rm crit,1}$; Eq.~\ref{eq:etacritslow}) are shown with solid purple lines. }
\label{fig:successful}
\end{figure}

Fig.~\ref{fig:successful} illustrates the parameter space of jet properties that give rise to successful versus unsuccessful jets in both the precessing (``misaligned'') and non-precessing (``aligned'') cases.  In non-precessing/aligned cases, even jets with very low efficiency $\eta \lesssim 10^{-4}$ and/or large opening angle $\theta_{\rm j} \gtrsim 0.1$ can be successful.  However, in the precessing/misaligned case, successful jets require either small misalignment angles $\psi$ or large efficiency. Interestingly, for sufficiently large misalignment angles $\psi \gtrsim 50^{\circ}$ (for $f_{\rm w} = 0.15; \beta_{\rm w} = 0.2$), even the most powerful jets with $\eta \approx 1$ are trapped. 

\section{Diverse Observable Signatures of Escaping Jets}
\label{sec:escape}   
In addition to the mass and radius of the disrupted star and the mass and spin of the SMBH, different TDEs will be characterized by different orbital penetration factors $\beta$ and initial disk-spin misalignment angles $\psi$. Furthermore, if the system produces a relativistic jet via the BZ mechanism, different quantities of magnetic flux threading the SMBH will lead to different jet efficiencies $\eta$ (Eq.~\eqref{eq:etaBZ}).  In this section we propose a unifying jetted TDE theory that explains why natural event-to-event variation in the parameters $\{ \beta, \eta, \psi\}$ gives rise to a wide range of observable jet signatures, ranging from the most powerful promptly escaping jets like J1644 to weaker jets which generated dimmer, delayed radio emission. 

\subsection{Minimum Luminosity/Energy of Successful Jets}
\label{5.1}

Fig.~\ref{fig:successful} shows that for typical jet misalignment angles $\psi \sim \mathcal{O}(1)$, a precessing TDE jet will initially be unsuccessful ($\eta < \eta_{\rm crit,2}$), unless the jet is very powerful (large $\eta$).  Because the conditions for a successful jet derived in Sec.~\ref{sec:jet head} are independent of time, an initially unsuccessful jet of fixed properties $\{\eta, \psi\}$ will remain unsuccessful at all later times.  However, as the jet becomes aligned with the SMBH spin axis  (over a timescale $t_{\rm align}$; Sec.~\ref{sec:alignment}), eventually the jet escape criterion $\eta > \eta_{\rm crit}$ (Eq.~\eqref{eq:etacritfast}) will be satisfied upon reaching some critical $\psi = \psi_{\rm crit}$ defined implicitly by $\eta = \eta_{\rm crit,2}(\psi_{\rm crit})$.  This critical condition defines the minimum power of the successful jet at the time of break-out:
\be
L_{\rm bo} = \eta_{\rm crit,2}(1-f_{\rm w})\dot{M}_{\rm fb}(t_{\rm align})c^{2},
\ee
and its corresponding isotropic luminosity,
\begin{eqnarray}
&& L_{\rm bo,iso} = \frac{2L_{\rm bo}}{\psi_{\rm crit}^{2}} \simeq \nonumber \\
&& f_{\rm w}\beta_{\rm w}(1-f_{\rm w})\left(1 + 4 \pi^2 \tan^2 \psi_{\rm crit} \right) \dot{M}_{\rm fb}(t_{\rm align})c^{2},
\label{eq:Lboiso}
\end{eqnarray}
where we have assumed the rapidly-precessing jet spreads its power over a solid angle $2\pi \psi^{2}$ corresponding to the entirety of the precession cone.  Likewise, the minimum total energy of an escaping jet can be approximated as
\begin{eqnarray}
E_{\rm bo} &=& \int_{t_{\rm bo}}^{\infty} L_{\rm j}dt \sim L_{\rm bo}t_{\rm align} \nonumber \\
&\simeq& 0.1\eta_{\rm crit}(1-f_{\rm w})M_{\star}c^{2}(t_{\rm align}/6t_{\rm fb})^{-2/3}.
\label{eq:Ebo}
\end{eqnarray}
Eq.~\eqref{eq:Ebo} can also be interpreted as the {\it maximum} energy deposited by a failed jet into the wind-envelope medium. 

\subsection{Distinct Classes of Jetted TDEs}

Table \ref{tab:1} summarizes three qualitatively different classes of jetted TDEs based on the timescale/mechanism of jet escape and the associated multi-wavelength emission properties.  

The first class of jets (first column in Tab.~\ref{tab:1}) are those characterized by high efficiencies ($\eta \geq \eta_{\rm crit,2}$; Eq.~\eqref{eq:etacritfast}) sufficient to escape the wind environment even prior to alignment of the jet axis with the SMBH spin on timescales $t_{\rm align} \lesssim t_{\rm fb}$.  Such jets will generally possess the highest isotropic luminosities carrying the largest energies $E_{\rm bo} \gtrsim 0.1\eta_{\rm crit,2}M_{\star}c^{2}$ (Eq.~\eqref{eq:Ebo}).  The early X-ray light curves of these prompt powerful jets will also likely exhibit signatures of precession because the jet has escaped prior to complete alignment.  Because the break-out can occur even at large $\psi = \psi_{\rm crit}$, the escaping jet can share its energy across a wide solid angle $\simeq \pi \psi^{2}$, larger than that of the intrinsic solid angle of the jet, $\pi \theta_{\rm j}^{2}$

The next two classes of TDE jets are those characterized by intermediate efficiencies $\eta_{\rm crit,1} \leq \eta \leq \eta_{\rm crit,2}$, for which jet escape occurs only once it has undergone nearly complete alignment on the timescale $t_{\rm align} \gtrsim t_{\rm fb}$. As both $\eta_{\rm crit,1}$ and $\eta_{\rm crit,2}$ depend on the accretion disk wind properties ($f_{\rm w}$, $\beta_{\rm w}$), TDEs with larger $f_{\rm w}$ or $\beta_{\rm w}$ require more powerful jets (higher $\eta$) to escape.  This category of jet escape can produce two distinct outcomes depending on the timescale of alignment (Classes 2 and 3, shown in the second and third columns in Tab.~\ref{tab:1}, respectively).  For relatively powerful jets with large $\eta$ or intermediate $\eta \gtrsim 10^{-3}-10^{-1}$ and high  $\beta$ (Fig.~\ref{fig:alignment}), the jet's strong magnetic field is again capable of aligning quickly via the magneto-spin mechanism ($t_{\rm align} \lesssim t_{\rm fb}$).  The combination of large jet efficiency $\eta$ and short alignment time, again imply that the luminosity and energy of the escaping jet are large. As in Class 1 events, the X-ray light curve will again exhibit a power-law decrease after alignment following the fall-back rate, but this time without a significant pre-alignment phase.  

By contrast, weaker jets with small $\eta \lesssim 10^{-3}$ or intermediate  efficiencies and small $\beta$  (Fig.~\ref{fig:alignment}) align slower through the hydrodynamic mechanism over timescales of months to years (Eqs.~\eqref{eq:talign1}).  In this case, the energy coupled to relativistic material can be much lower $E_{\rm bo} \ll 0.1\eta M_{\star}c^{2} \lesssim 10^{-4}M_{\odot}c^{2} \sim 10^{50}$ erg, and the effective opening angle of the precessing jet at the time of break-out will be smaller $\psi_{\rm crit} \lesssim 0.1.$  Such weak jets, which emerge substantially delayed from the time of disruption, will create late-rising radio afterglows, similar to those of the late-rising radio flares observed following TDE \citep{Horesh+21,Horesh+21b,Perlman+22,Sfaradi+22,Cendes+22}.
This class of jets may also produce internal X-ray emission, though the much lower accretion rates at such late times will likely render the X-ray emission mechanism different from those of promptly escaping jets created during highly-super Eddington phase.  The afterglow produced by interaction of the jet with the external medium will also produce transient X-ray emission, which may also manifest as a hard spectral component on top of the soft quiescent disk emission.

\begin{deluxetable*}{|c|c|c|c|}
\tablewidth{700pt}
\tabletypesize{\scriptsize}
\tablehead{
\colhead{\textbf{Class}} & \colhead{ \textbf{1.~Jet escaping prior to alignment}} & 
\colhead{\textbf{2.~Jet escaping at alignment} }  & \colhead{\textbf{3.~Delayed escaping mildly relativistic jet}}
} 
\startdata
\hline 
Jet efficiency criterion & $\eta \geq \eta_{\rm crit,2} ( \psi, \beta_{\rm w},f_{\rm w})$&  $\eta_{\rm crit,1} \leq \eta \leq \eta_{\rm crit,2}$  &  $\eta_{\rm crit,1} \leq \eta \leq \eta_{\rm crit,2}$ \\ 
\hline
Alignment mechanism $(\eta, \beta)$ & \multicolumn{2}{c|}{Magneto-spin alignment}  &  Hydrodynamic alignment \\ 
\hline 
 X-ray Properties & Luminous non-thermal emission; & Luminous non-thermal emission; & Transient X-ray afterglow emission; \\ 
 & Precession signatures as jet aligns; & Power-law decay after alignment &  \\
 & Power-law decay after alignment & & \\
 \hline
X-ray Light Curve  & & & \\
& \includegraphics[width=0.23\textwidth]{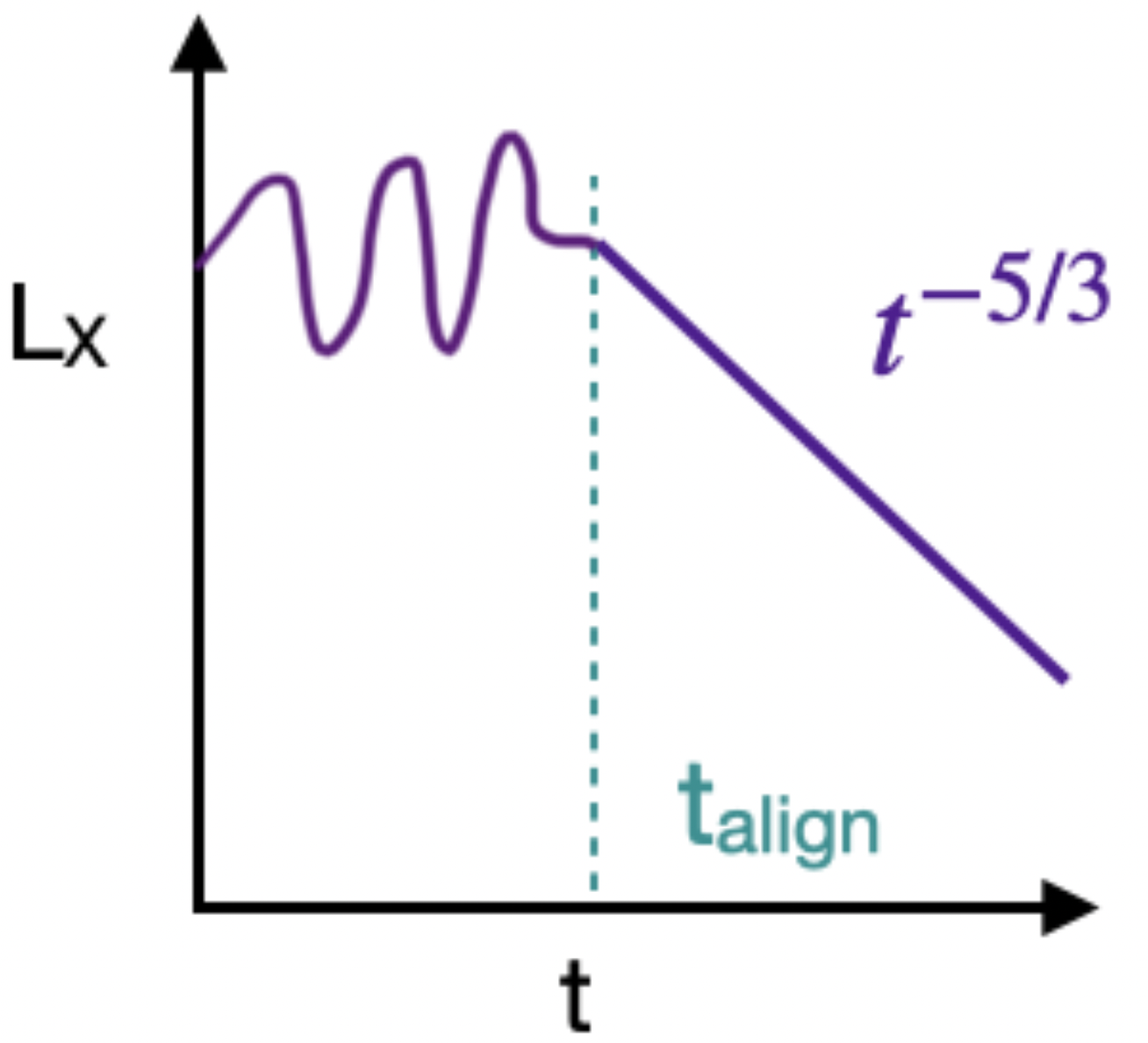} & \includegraphics[width=0.23\textwidth]{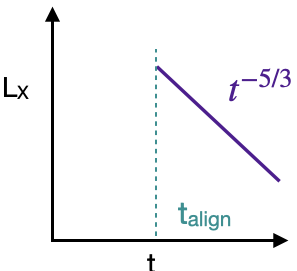} & \\
\hline
 Radio Properties   & \multicolumn{2}{c|} {Prompt afterglow emission from ultra-relativistic jet}  & \textbf{Delayed} afterglow from transrelativistic jet \\
 \hline
   Candidate Events & J1644 & J2058; J1112 & ASASSN-15oi, iPTF16fnl, IGRJ12580+013, \\
   &  & AT2022cmc & AT2018hyz, AT2019azh\\
\enddata
\caption{ 
\label{tab:1} Diversity of outcomes of jetted TDEs depending on the initial misalignment angle $\psi$, jet efficiency $\eta$, orbital penetration factor $\beta$ and the physical mechanism responsible for alignment (Sec.~\ref{sec:alignment}).  The jet alignment mechanism depends on both the jet efficiency $\eta$ and orbital penetration factor $\beta$ (Fig.~\ref{fig:alignment}).  The jet-spin misalignment angle $\psi$ and wind properties $f_{\rm w}$ and $\beta_{\rm w}$ in part determine the critical efficiency $\eta_{\rm crit,2}$  (Eq.~\eqref{eq:etacritfast}) needed to escape before alignment  and after alignment  $\eta_{\rm crit,1}$ (Eq.~\eqref{eq:etacritslow}) and hence on the energy.}
\end{deluxetable*}

\subsection{Comparison to Jetted TDE Sample}

Four powerful relativistic on-axis jetted TDE flares have been observed to date \citep{Andreoni22,Bloom+11,Cenko12,Brown15}, while late-rising radio emission delayed by months to years has been observed from several optically-selected TDE (\citealt{Horesh+21,Horesh+21b, Perlman+22,Sfaradi+22}).  In this section, we provide an exemplar TDE from each of the three jetted TDE classes in the unification scenario summarized in Table \ref{tab:1} and model its synchrotron radio afterglow.

\subsubsection{Swift J1664 as a jet escaping prior to alignment}

\paragraph{X-ray Emission.}
X-ray observations of J1644 (e.g., \citealt{Bloom+11,Levan+11,Burrows+11}) reveal three distinct emission phases:\\
\begin{enumerate}
    \item At early times after the X-ray trigger ($t \lesssim 30$ d), the internal jet emission is highly time-variable, consistent with the jet opening angle entering and exiting the observer line of sight, similar to the expected behavior for a wobbling and precessing jet.
\item At intermediate times (30 d $\lesssim t \lesssim 450$ d) the X-ray light curve decays at a rate consistent with the expected rate of fall-back accretion, $L_{\rm X} \propto t^{-5/3}$.
\item At late times ($t \approx 500$ d) the X-ray luminosity abruptly drops by several orders of magnitude (e.g., \citealt{Zauderer+11}), likely when the disk transitions to a thin-disk state around $t \sim t_{\rm Edd}$ (Eq.~\eqref{eq:tEdd}) and the SMBH loses its magnetic flux (e.g.,~\citealt{Tchekhovskoy+14}). 
\end{enumerate}

We propose J1644 as an exemplar TDE that produces a relativistic jet which escapes from the ejecta prior to alignment (first column in Tab.~\ref{tab:1}).  The phase of jet alignment ($t_{\rm align} \approx 30$ d), corresponds to the transition between Stages 1 and 2 above.  Such a short alignment time favors the magneto-spin alignment mechanism (Fig.~\ref{fig:alignment}), consistent with a high-efficiency Poynting-flux dominated ultra-relativistic jet.  Prior to alignment ($t \leq 30$ d) the emission seen by the observer comes from a structure with a large half-opening angle $\psi$, corresponding to the initial misalignment angle between the SMBH spin axis and the jet.  The fact that  X-ray emission is detectable prior to alignment favors a relatively modest misalignment angle (for larger misalignment angles, emission beamed along the spin axis would be harder to observe). 

After alignment completes at $t \gtrsim 30$ d, the observed luminosity follows $L_{j}|_{t_{\rm align}}(t+t_{\rm align})^{-5/3}$, with the remaining jet energy $E_{\rm bo}$ (Eq.~\eqref{eq:Ebo}) focused in a narrow jet of half-opening angle $\theta_{\rm j}$, until the jet shuts off at $t \approx 500$ d.  In addition to this core jet, the larger-angle structured jet emitted during the precession (phase 1) above, also contributes to the emission, resulting in a two-component jet. A jet escaping prior to alignment thus naturally explains the two-component structure for J1644, which \citet{Mimica+15} also found was needed to explain the radio observations.  

\paragraph{Radio Emission.} 
We model the radio emission from each jet of the two-component structure as synchrotron radiation from relativistic electrons accelerated behind the shocks created as the relativistic jet collides with the circumnuclear medium using a modified version of our semi-analytical model presented in \citet{Odelia1}. 

The early-time radio data across a range of frequencies $4.9-24$ GHz \citep{Zauderer+11} is well fitted by a power law $F_\nu \propto t^\alpha $ with $\alpha \approx$ 2, followed by a shallower increase with $\alpha \approx$ 0.5 around $t \approx t_{\rm RS} \approx 10$ d. This transition at $t_{\rm RS}$ is achromatic and well explained by a transition from the early phase when the reverse shock crosses the ejecta shell, to the classical adiabatic blast wave expansion at later times \citep{Metzger+12}.  Based on the early phase emission, the inferred radial density profile of the circumnuclear medium (CNM) was consistent with a power-law of the form $\rho_{\rm ext} = A r^{-2}$, similar to that expected for a steady wind or Bondi accretion onto the SMBH ($\rho_{\rm ext} \propto r^{-3/2}$).  Accordingly, we model the early light curve evolution following Eqs.~(7)-(12) of \citet{Metzger+12} for a wind-type medium.  Then, after the reverse shock crossing time $t_{\rm RS}$, the ultra-relativistic forward shock is assumed to follow the standard \citet{Blandford&McKee76} adiabatic self-similar evolution, before finally transitioning (once the shock Lorentz factor reaches $\Gamma \approx \sqrt{2}$) to the Newtonian Sedov-Taylor evolution at late times.  We make the standard assumption that fixed fractions $\epsilon_B $ and $\epsilon_e$ of the internal energy of the post-shock gas goes into the energy of the magnetic field and relativistic electrons, respectively; the latter are assumed to acquire a power law distribution of energy, immediately behind the shock $N(\gamma_{\color{black}e}) \propto \gamma_{\color{black}e} ^{-p}$.   We calculate the synchrotron spectrum following \citet{Granot&Sari02}.

After $t \approx 30-35$ d the light-curve power-law slightly increases from $\alpha \approx$ 0.5 to $\alpha \approx$ 0.7 up to the peak time, which depends on the observing frequency.  After the peak, there are two phases: a shallow decline with $\alpha \approx -$(0.3-0.6) depending on radio frequency, followed by a sharper decay $\alpha \approx -$(1.7-2). Here we interpret this late behavior as a transition to the non-relativistic phase. Indeed, for $\nu \geq \nu_m$, the flux $F_\nu$ is predicted to steepen after the non-relativistic to $F_\nu \propto t^{(5-7p)/6}$ \citep{Chevalier}, consistent with the late-time decay for $p \approx 2.4.$  

\begin{figure}[h!]
 	\centerline{\includegraphics[width=90mm]{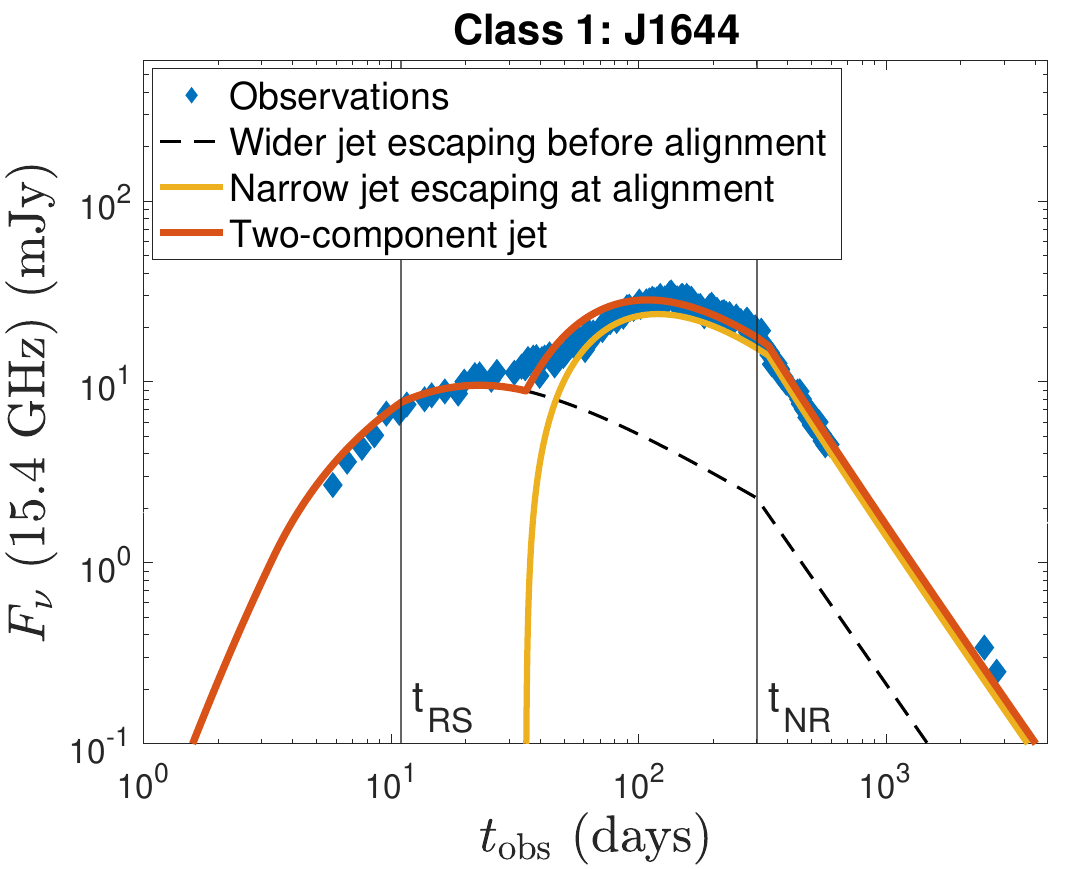}}
    \caption{Two-component synchrotron emission model fit to the 15.4 GHz radio light curve of Swift J1644+57 as a relativistic jet that escapes from the TDE wind prior to alignment with the SMBH spin into an external medium with a wind-like density profile $\rho _{\rm ext}= A r^{-2}$, where $A = 3\times 10^{11}$ g cm$^{-1}$.  Prior to alignment, precession of the jet creates a large angular structure with $E_{\rm iso}=10^{53}$ erg and half-opening angle $\psi =20^{\circ}$ corresponding to the misalignment angle.  After alignment completes, the remainder of the jet energy is injected into a narrow jet with $E_{\rm iso}=6 \times 10^{52}$ erg and half-opening angle $\theta_{\rm j} =5^{\circ}$ aligned with the SMBH spin axis and viewer line of site. The two vertical lines depict the dynamical times of interest: $t_{\rm RS}$, the shell crossing time and $t_{\rm NR}$, the time where the blast wave transitions to the non-relativistic Newtonian regime.  Observations data are taken from \citep{Zauderer+11,Berger+12, Eftekhari}.  We assume $p=2.4$ and $\epsilon_e = \epsilon_B = 0.1$ for the electron power law and shock equipartition parameters, respectively.}
   \label{fig:J1644}
\end{figure}

\subsubsection{AT2022cmc as a Jet Escaping at Alignment}
\paragraph{X-ray observations}
Unlike J1644, the X-ray observations of AT2022cmc peak over just a few days and are consistent with decaying as a power-law thereafter $L_{\rm X} \propto t^{-\alpha}$ with $\alpha \approx 2$ from the beginning \citep{Andreoni22}. We propose AT2022cmc as an exemplar TDE whose relatively powerful jet escaped from the ejecta relatively promptly only at the point of final alignment (second column in Tab.~\ref{tab:1}).  The relatively quick alignment time in this case again points towards rapid magneto-spin alignment (Sec.~ \ref{sec:alignment}), consistent with the high efficiency $\eta$ of a powerful jet.
\paragraph{Radio Emission}
\citet{Andreoni22} present radio detections between $5.1-45.3$ days across a range of frequencies $5-350$ GHz. The spectrum is well fit by a broken power law $F_\nu \propto \nu^\beta $ with $\beta \approx$ 2 at low frequencies $ \nu \leq \nu_a$, $\beta \approx$ 1/3 at intermediate frequencies $ \nu_a \leq \nu \leq \nu_m $ and $\beta \approx -1$ at high frequencies $ \nu_m \leq \nu$.  This is consistent with synchroton emission of spectral type 1 from \cite{Granot&Sari02} for $p \sim 2.9$.  We note that emission was also detected at $15.5$ GHz from $14$ days to $90$ days by AMI-LA \citep{Rhodes}. This emission exhibited a non-negligible level of variability; if intrinsic (e.g. not due to scintillation), which may suggest an ``internal'' contribution to the jet emission in addition to that of an external shock (e.g., \citealt{vanVelzen+13}).

As in the case above, we model the emission as the synchrotron radio afterglow from relativistic electrons accelerated behind the shocks created as the relativistic jet collides with the circumnuclear medium using a modified version of our semi-analytical model discussed in \citet{Odelia1}, this time for only a single jet component.  

\begin{figure}[h!]
 	\centerline{\includegraphics[width=90mm]{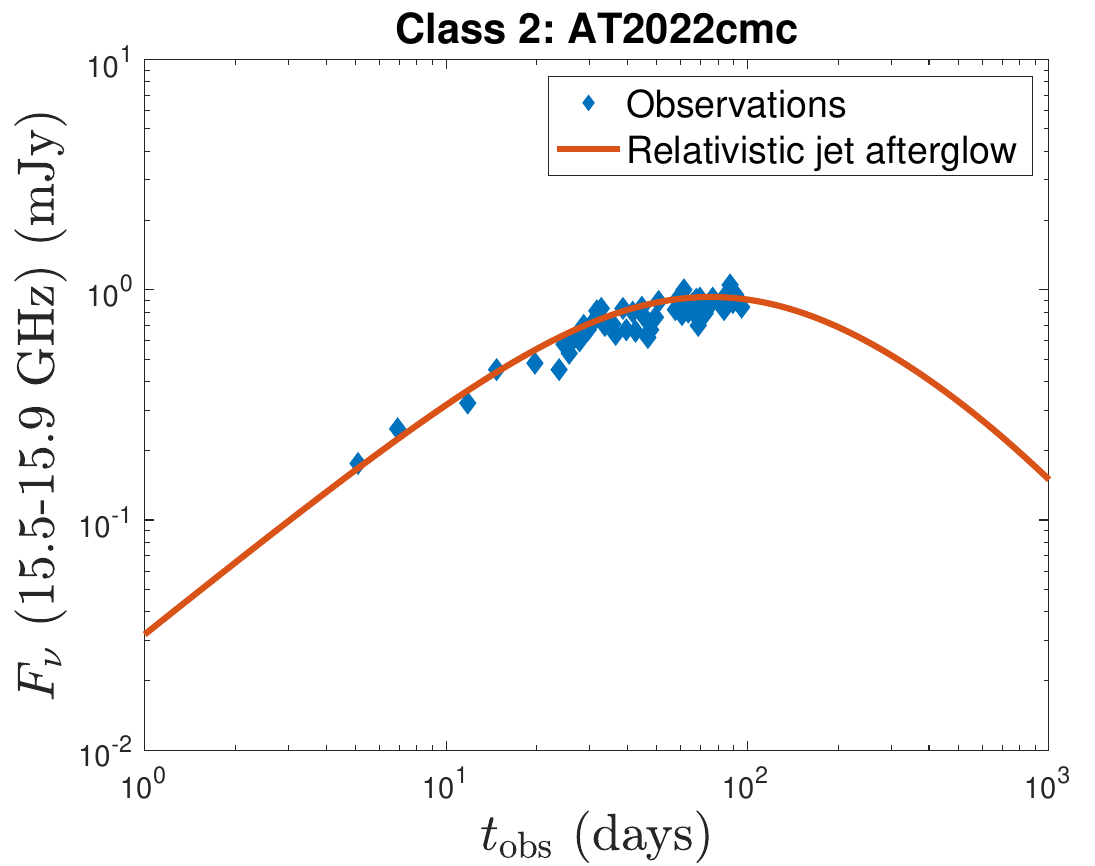}}
    \caption{Synchrotron emission model of AT2022cmc as a relativistic jet that escapes from the TDE wind at the time of alignment with the SMBH spin computed with the model described in \cite{Odelia1}.
    The relativistic jet has an isotropic energy  $E_{\rm iso}=1 \times 10^{53}$ erg and half-opening angle $\theta_{\rm j} =10^{\circ}$. The external medium harbors  wind-like density profile $\rho _{\rm ext}= A r^{-2}$, where $A = 3\times 10^{12}$ g cm$^{-1}$. Observations data are taken from \citep{Andreoni22,Rhodes}.  We assume $p=2.9$, $\epsilon_e =0.2 $ and $ \epsilon_B = 0.002$ for the electron power law and shock equipartition parameters, respectively.}
   \label{fig:AT2022}
\end{figure}

\subsection{AT2018hyz as a Delayed Escaping Mildly Relativistic Jet}
\paragraph{X-ray observations}

The X-ray light curve of AT2018hyz detected by {\it Swift} XRT exhibited a flat evolution \citep{Gomez20}, its flux at $t=1253$ days after optical maximum fading only factor of 2 compared that at 86 days \citep{Cendes+22}.  Though the time coverage was not dense, no abrupt change in the X-ray properties were observed around the onset time of the radio flare we shall now discuss. 
\paragraph{Radio emission}

AT2018hyz was initially undetected at radio frequencies, showing only upper limits at times from 32 days ($< 85 \mu$ Jy at 15.5 GHz; \citealt{Horesh18}) until 705 days ($< 0.45$ mJy at 3 Ghz; \citealt{Lacy20}) after optical discovery (along with a few other upper limits reported between 32 days and 705 days; \citealt{Gomez20,Murphy21,Lacy20}).  However, from 972 days until 1296 days, AT20a8hyz began to brighten considerably, showing detectable radio emission across a range of frequencies $0.89-240$ GHz \citep{Cendes+22}.  The spectrum is well fit by a broken power-law $F_\nu \propto \nu^\beta $ with $\beta \approx -0.65$ at higher frequencies $\nu_p \leq \nu$, but with the spectral shape less well-constrained for lower frequencies $\nu \leq \nu_p$, consistent with a synchroton spectrum type 2 of \cite{Granot&Sari02} with $p \sim 2.25$.  However, the steep rise in the radio flux $F_{\nu} \propto t^\alpha$ with $\alpha \approx 4.8-6$ (depending on frequency $\nu$) was unexpected in an on-axis afterglow interpretation, at least provided that $t$ is measured with respect to the optical peak.  

Though puzzling in comparison to powerful jetted TDEs like J1644, the observed features of AT2018hyz are consistent with a delayed escaping weaker jet within the framework of our unified theory of jetted TDE (final column in Tab.~\ref{tab:1}).  The apparently steep-rising radio light curve can be understood if the zero-point of the afterglow emission is not the time of the disruption or onset of the optical emission, but rather the time when the jet finally aligns sufficiently with the SMBH spin axis to escape from the TDE environment. Relatively weak jets align through the hydrodynamic mechanism (Sec.~\ref{sec:alignment}) which can occur after several months or years (Eq.~\eqref{eq:talign1}), consistent with the timing of the first radio detection from AT2018hyz. As discussed in Sec.~\ref{5.1}, because of the lower efficiency and accretion rate associated with such a delayed escaping jet, it can possess a lower characteristic energy $E \sim 10^{50}$ erg, causing it to decelerate rapidly and expand into the circumnuclear medium at transrelativistic speeds.  

The lack of bright jetted X-ray emission from AT2018hyz is also consistent with this scenario: because the jet is relatively weak, its transient afterglow emission can be outshone by the accretion disk at these late times.  As an application of this scenario, we model in Fig.~\ref{fig:AT2022} the radio observations of AT2018hyz as the synchrotron afterglow emission from a delayed on-axis jet, which we assume escaped the ejecta and began to interact with the circumnuclear medium at $t_{\rm align} \approx 920$ days after the optical discovery.  The low jet energy we derive $E \sim 10^{50}$ erg is consistent with that expected in the delayed escaping jet scenario.   

\begin{figure}[h!]
 	\centerline{\includegraphics[width=90mm]{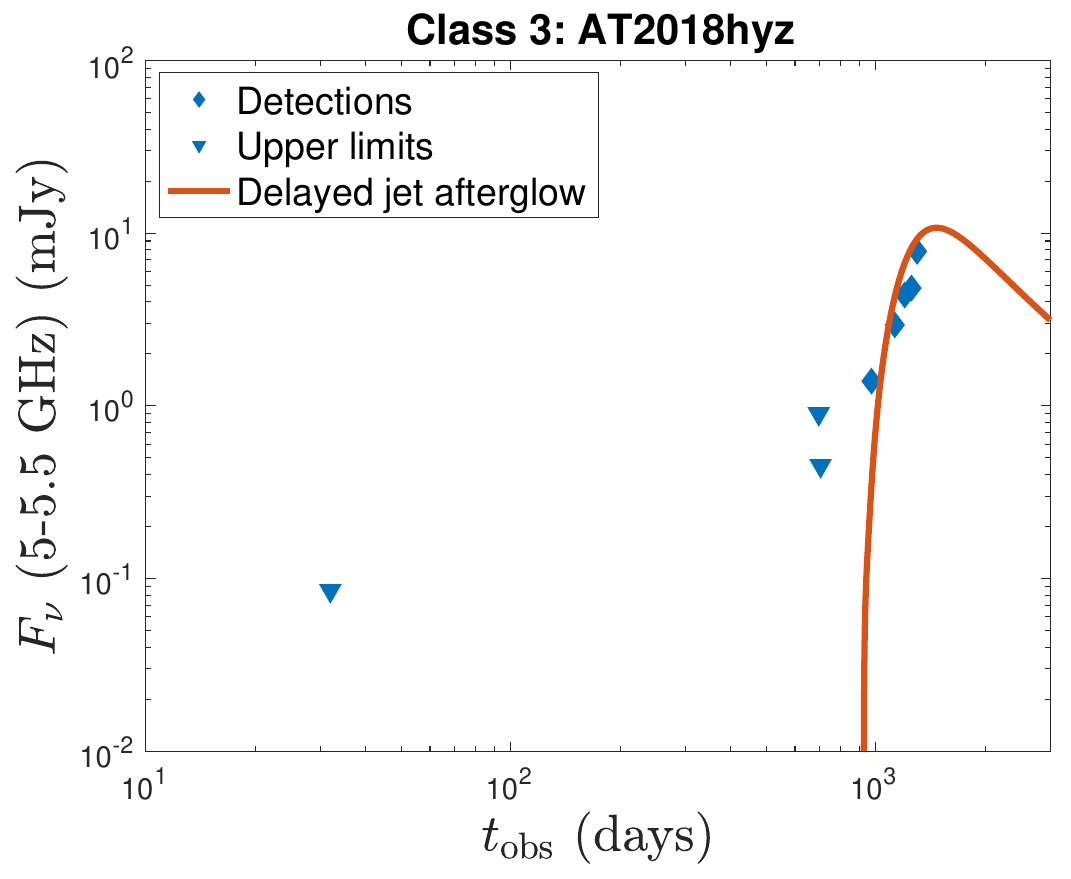}}
    \caption{Synchrotron emission model of AT2018hyz as a delayed mildly relativistic jet that escapes at the time of alignment $t_{\rm align} \sim 920$ days with the SMBH spin computed with the model described in \cite{Odelia1}.
    The jet has an isotropic energy  $E_{\rm iso}=8 \times 10^{49}$ erg and half-opening angle $\theta_{\rm j} =20^{\circ}$.  As in the other jetted TDE, we assume the external medium possesses a wind-like density profile $\rho _{\rm ext}= A r^{-2}$, where $A = 9\times 10^{12}$ g cm$^{-1}$. Observations data are taken from \citep{Cendes+22} and upper limits are taken from \citep{Horesh18, Lacy20}. We assume $p=2.25$, $\epsilon_e =0.2 $ and $ \epsilon_B = 0.007$ for the electron power law and shock equipartition parameters, respectively.}
   \label{fig:AT2018}
\end{figure}


\section{Conclusions}
\label{sec:conclusions}

We have proposed a unified model for jet production in TDEs that can account for both promptly-escaping powerful ultra-relativistic jets as well as weaker delayed-escaping mildly-relativistic jets.  In our scenario, these differences in outcome depend primarily on event-to-event variations in the intrinsic TDE properties (initial orbit-spin misalignment angle $\psi$, orbital penetration factor $\beta$, jet efficiency $\eta \equiv L_{\rm j}/\dot{M}c^{2}$).

Assuming the orbital planes of the victim stars are distributed isotropically with respect to the spin axis of the SMBH, most TDE accretion disks will form with large misalignment angles relative to the narrow jet opening angle.  Analytic estimates and GRMD simulations show that such misaligned disk/jet systems precess together due to the Lense-Thirring effect \citep{Liska+18,Chatterjee20}. We estimate analytically the time-dependent precession period of a thick TDE disk, taking into account: i) the innermost radius (ISSO) of a misaligned disk; ii) the time-dependent outer radius due to viscous spreading, accounting for the possibility of wind angular momentum loss; iii) the addition of mass to the disk from TDE fallback accretion (Sec.~\ref{sec:disk/jet}, Fig.\ref{fig:Tprec}).  

For the mid- to high-levels of black hole spin required to launch a jet, we find typical precession periods of order a few days. Although most jets begin misaligned, two families of mechanisms can act to align the disk/jet with the SMBH spin axis over time: magneto-spin and hydrodynamic alignments. We estimate the typical alignment timescale to be weeks for the magneto-spin mechanism and months to years for the hydrodynamic mechanisms.  We showed that the dominant alignment mechanism depends on the orbital penetration factor $\beta$ and Blandford-Znajek jet efficiency $\eta$ (Fig.~\ref{fig:alignment}). Powerful jets with $\eta \sim 0.1 - 1$ undergo magneto-spin alignment while weak jets with $\eta \leq 10^{-3}$ align through the hydrodynamic mechanisms. Jets with intermediate efficiencies can align through either mechanism, depending on $\beta$. These two very different timescales give rise to a dichotomy of jet escape times and associated jet signatures.  
\\

As $t_{\rm prec} \ll t_{\rm align}$, misaligned jets likely undergo a large number of precession cycles before alignment.  Such rapid precession also implies that outflows from the disk will encase the TDE system in a quasi-spherical wind envelope on large scales (Fig.~\ref{fig:cartoon}; Sec.~\ref{sec:wind}), which determines the gaseous environment from which a TDE jet must escape to generate observable X-ray or radio emission.  We derived a critical jet efficiency $\eta$ required for aligned and misaligned/precessing jets to escape from the disk-wind ejecta (Sec.~\ref{sec:jet head}).  We showed that the conditions for a misaligned jet undergoing rapid precession to escape are far more stringent than for an aligned jet with a fixed drilling direction (Fig.~\ref{fig:successful}).

The critical jet escape efficiency defines a minimum energy injected by the jet at the break-out time $E_{\rm bo}$ (Eq.~\eqref{eq:Ebo}).  All else being equal, jets which require longer to align and break out will inject less energy into the external circumnuclear medium once they do so.  Powerful relativistic jets that escape promptly will have $E_{\rm iso} \sim 10^{52-54}$ erg, while delayed mildly relativistic jet will typically have $E_{\rm iso} \sim 10^{48-50}$ erg.  Depending on the intrinsic properties of the system $\{\psi, \beta$, $\eta$, $\beta_w$, $f_w\}$, we propose that every jetted TDE belongs to one of three classes (Tab.~\ref{tab:1}):
\begin{enumerate}
   \item Relativistic jet which escapes prior to alignment 
    \item Relativistic jet which escapes at alignment
   \item Delayed mildly relativistic jet
\end{enumerate}
As an explicit demonstration, we model the jetted radio afterglow emission of an exemplar TDE from each class:  J1644 as a relativistic jet escaping before alignment (Fig.~\ref{fig:J1644}), AT2022cmc as a relativistic jet escaping at alignment (Fig.~\ref{fig:AT2022}) and AT2018hyz as a delayed mildly relativistic emission (Fig.~\ref{fig:AT2018}), using an adapted version of our semi-analytical model \citep{Odelia1}.  As expected, going from Class 1 to Class 3, the jet energy required by the afterglow model decreases accordingly. In addition, the X-ray observations of each exemplar TDE are consistent with this theoretical picture: J1644 X-ray light curve exhibited precession signatures followed by a power-law decay, AT2022cmc showed an X-ray power-law decay and AT2018hyz did not exhibit signatures of a relativistic jet. 

 The four relativistic jetted TDE were all observed at high redshifts $z \approx 0.35- 1.2$, while those optically-selected TDE showing delayed radio flares are much closer $z \sim 0.004 -0.07$.  This suggests that the fraction of TDEs which are accompanied by weak delayed radio flares greatly exceeds those which produce prompt powerful jets.  Indeed, given an initially isotropic distribution of inclination $\psi$ and a wide range of jet efficiencies $\eta$ (e.g., defined by a range of SMBH spin magnitudes and/or magnetic flux), we should expect delayed weak jets (Class 3) to be far more common than promptly escaping powerful jets (Class 1).  If the magnetic flux threading the SMBH originates exclusively from the original magnetic field of the disrupted star, values of $\eta \ll 10^{-3}$ are predicted for realistic internal stellar magnetic fields \citep{Giannios&Metzger11}. By contrast, if magnetic field amplification occurs during circularization \citep{Bonnerot+17} or later in the accretion disk, or enough pre-existing magnetic flux is present in the galactic nucleus, efficiencies up to the MAD limit $\eta \sim 1$ may be possible (e.g., \citealt{Tchekhovskoy+14}).

Given this wide range of possible $\eta$, our scenario predicts a larger number of delayed radio flares with energy $10^{48}-10^{50}$ erg than promptly escaping powerful jets. Moreover, X-ray observations around the time of jet escape would provide an important constraint on our delayed jet model. As jets which undergo late escape are necessarily weak in our scenario, their bolometric luminosity, or that of their X-ray afterglow, is likely to be outshone by the thermal emission of the accretion disk. Even if their transient X-ray afterglow is observed it could be hard to disentangle it from the non-thermal disk spectrum which develops following a state transition (as observed in X-ray binaries; \citealt{Fender+04}), which provides an alternative model for delayed radio jets in TDEs. However, a clear signature to distinguish our scenario from a state transition would be the presence of a continuous spectrum. Indeed, the afterglow scenario predicts a single synchrotron spectrum spanning radio to non-thermal X-ray frequencies, while in a state transition scenario, the radio and X-ray emission mechanisms are distinct. 

In principle, TDEs can serve as laboratories for studying the properties of jet formation in ways complementary to those in AGN. It remains debated why $\sim 10\%$ of AGN are radio loud, and whether this is related to an intrinsic property of the black hole such as its spin (e.g., \citealt{Tchekhovskoy+10b}) or an external property of the external environment like the magnetic flux (e.g., \citealt{Kelley+14}).  If radio-loud AGNs are related to high BH spin, this would, at least naively, predict that the jetted fraction of AGN would match that of TDEs, assuming both sample similar SMBH properties.  The higher intrinsic TDE jet launching fraction predicted by our model would alleviate the discrepancy between the observed $\sim 1\%$ jetted TDE fraction and the $\sim 10\%$ radio bright AGN fraction.

\acknowledgements
The authors would like to thank N.~Stone and E.~Quataert for helpful discussions. OT gratefully acknowledges the support of the Einstein-Kaye scholarship and of the Israel Ministry of Science and Technology.  BDM was supported in part by the NSF through the NSF-BSF program (grant number AST-2009255).  The Flatiron Institute is supported by the Simons Foundation.   

\bibliography{ms}

\begin{thebibliography}{}
\expandafter\ifx\csname natexlab\endcsname\relax\def\natexlab#1{#1}\fi
\providecommand{\url}[1]{\href{#1}{#1}}
\providecommand{\dodoi}[1]{doi:~\href{http://doi.org/#1}{\nolinkurl{#1}}}
\providecommand{\doeprint}[1]{\href{http://ascl.net/#1}{\nolinkurl{http://ascl.net/#1}}}
\providecommand{\doarXiv}[1]{\href{https://arxiv.org/abs/#1}{\nolinkurl{https://arxiv.org/abs/#1}}}

\bibitem[{{Alexander} {et~al.}(2016){Alexander}, {Berger}, {Guillochon},
  {Zauderer}, \& {Williams}}]{Alexander16}
{Alexander}, K.~D., {Berger}, E., {Guillochon}, J., {Zauderer}, B.~A., \&
  {Williams}, P.~K.~G. 2016, \apjl, 819, L25,
  \dodoi{10.3847/2041-8205/819/2/L25}

\bibitem[{{Alexander} {et~al.}(2020){Alexander}, {van Velzen}, {Horesh}, \&
  {Zauderer}}]{Alexander+20}
{Alexander}, K.~D., {van Velzen}, S., {Horesh}, A., \& {Zauderer}, B.~A. 2020,
  \ssr, 216, 81, \dodoi{10.1007/s11214-020-00702-w}

\bibitem[{{Alexander} {et~al.}(2017){Alexander}, {Wieringa}, {Berger},
  {Saxton}, \& {Komossa}}]{Alexander+17}
{Alexander}, K.~D., {Wieringa}, M.~H., {Berger}, E., {Saxton}, R.~D., \&
  {Komossa}, S. 2017, \apj, 837, 153, \dodoi{10.3847/1538-4357/aa6192}

\bibitem[{{Andreoni} {et~al.}(2022){Andreoni}, {Coughlin}, {Perley}, {Yao},
  {Lu}, {Cenko}, {Kumar}, {Anand}, {Ho}, {Kasliwal}, {de Ugarte Postigo},
  {Sagu{e}s-Carracedo}, {Schulze}, {Kann}, {Kulkarni}, {Sollerman}, {Tanvir},
  {Rest}, {Izzo}, {Somalwar}, {Kaplan}, {Ahumada}, {Anupama}, {Auchettl},
  {Barway}, {Bellm}, {Bhalerao}, {Bloom}, {Bremer}, {Bulla}, {Burns},
  {Campana}, {Chandra}, {Charalampopoulos}, {Cooke}, {D'Elia}, {Das}, {Dobie},
  {Agu}~Fernandez, {Freeburn}, \& {Fremling}}]{Andreoni22}
{Andreoni}, I., {Coughlin}, M.~W., {Perley}, D.~A., {et~al.} 2022, \nat, 612,
  430, \dodoi{10.1038/s41586-022-05465-8}

\bibitem[{{Bardeen} \& {Petterson}(1975)}]{Bardeen&Petterson75}
{Bardeen}, J.~M., \& {Petterson}, J.~A. 1975, \apjl, 195, L65,
  \dodoi{10.1086/181711}

\bibitem[{{Bate} {et~al.}(2000){Bate}, {Bonnell}, {Clarke}, {Lubow}, {Ogilvie},
  {Pringle}, \& {Tout}}]{Bate+00}
{Bate}, M.~R., {Bonnell}, I.~A., {Clarke}, C.~J., {et~al.} 2000, \mnras, 317,
  773, \dodoi{10.1046/j.1365-8711.2000.03648.x}

\bibitem[{{Begelman}(1979)}]{Begelman79}
{Begelman}, M.~C. 1979, \mnras, 187, 237, \dodoi{10.1093/mnras/187.2.237}

\bibitem[{{Berger} {et~al.}(2012{\natexlab{a}}){Berger}, {Zauderer}, {Pooley},
  {Soderberg}, {Sari}, {Brunthaler}, \& {Bietenholz}}]{Berger12}
{Berger}, E., {Zauderer}, A., {Pooley}, G.~G., {et~al.} 2012{\natexlab{a}},
  \apj, 748, 36, \dodoi{10.1088/0004-637X/748/1/36}

\bibitem[{{Berger} {et~al.}(2012{\natexlab{b}}){Berger}, {Zauderer}, {Pooley},
  {Soderberg}, {Sari}, {Brunthaler}, \& {Bietenholz}}]{Berger+12}
---. 2012{\natexlab{b}}, \apj, 748, 36, \dodoi{10.1088/0004-637X/748/1/36}

\bibitem[{{Blandford} \& {McKee}(1976)}]{Blandford&McKee76}
{Blandford}, R.~D., \& {McKee}, C.~F. 1976, Physics of Fluids, 19, 1130,
  \dodoi{10.1063/1.861619}

\bibitem[{{Blandford} \& {Znajek}(1977)}]{Blandford&Znajek77}
{Blandford}, R.~D., \& {Znajek}, R.~L. 1977, \mnras, 179, 433,
  \dodoi{10.1093/mnras/179.3.433}

\bibitem[{{Bloom} {et~al.}(2011){Bloom}, {Giannios}, {Metzger},
  {et~al.}}]{Bloom+11}
{Bloom}, J.~S., {Giannios}, D., {Metzger}, B.~D., {et~al.} 2011, Science, 333,
  203, \dodoi{10.1126/science.1207150}

\bibitem[{{Bollimpalli} {et~al.}(2023){Bollimpalli}, {Fragile}, \&
  {Klu{\'z}niak}}]{Bollimpalli+23}
{Bollimpalli}, D.~A., {Fragile}, P.~C., \& {Klu{\'z}niak}, W. 2023, \mnras,
  520, L79, \dodoi{10.1093/mnrasl/slac155}

\bibitem[{{Bonnerot} {et~al.}(2017){Bonnerot}, {Rossi}, \&
  {Lodato}}]{Bonnerot+17}
{Bonnerot}, C., {Rossi}, E.~M., \& {Lodato}, G. 2017, \mnras, 464, 2816,
  \dodoi{10.1093/mnras/stw2547}

\bibitem[{{Bonnerot} \& {Stone}(2021)}]{Bonnerot&Stone21}
{Bonnerot}, C., \& {Stone}, N.~C. 2021, \ssr, 217, 16,
  \dodoi{10.1007/s11214-020-00789-1}

\bibitem[{{Bower} {et~al.}(2013){Bower}, {Metzger}, {Cenko}, {Silverman}, \&
  {Bloom}}]{Bower+13}
{Bower}, G.~C., {Metzger}, B.~D., {Cenko}, S.~B., {Silverman}, J.~M., \&
  {Bloom}, J.~S. 2013, \apj, 763, 84, \dodoi{10.1088/0004-637X/763/2/84}

\bibitem[{{Bradnick} {et~al.}(2017){Bradnick}, {Mandel}, \&
  {Levin}}]{Bradnick+17}
{Bradnick}, B., {Mandel}, I., \& {Levin}, Y. 2017, \mnras, 469, 2042,
  \dodoi{10.1093/mnras/stx1007}

\bibitem[{{Brown} {et~al.}(2015){Brown}, {Levan}, {Stanway}, {Tanvir}, {Cenko},
  {Berger}, {Chornock}, \& {Cucchiaria}}]{Brown15}
{Brown}, G.~C., {Levan}, A.~J., {Stanway}, E.~R., {et~al.} 2015, \mnras, 452,
  4297, \dodoi{10.1093/mnras/stv1520}

\bibitem[{{Brown} {et~al.}(2017){Brown}, {Levan}, {Stanway}, {Kr{\"u}hler},
  {Tanvir}, {Davies}, {Fruchter}, {Cenko}, \& {Metzger}}]{Brown17}
---. 2017, \mnras, 472, 4469, \dodoi{10.1093/mnras/stx2193}

\bibitem[{{Burrows} {et~al.}(2011)}]{Burrows+11}
{Burrows}, D.~N., {et~al.} 2011, \nat, 476, 421, \dodoi{10.1038/nature10374}

\bibitem[{{Cendes} {et~al.}(2022){Cendes}, {Berger}, {Alexander}, {Gomez},
  {Hajela}, {Chornock}, {Laskar}, {Margutti}, {Metzger}, {Bietenholz},
  {Brethauer}, \& {Wieringa}}]{Cendes+22}
{Cendes}, Y., {Berger}, E., {Alexander}, K., {et~al.} 2022, arXiv e-prints,
  arXiv:2206.14297.
\newblock \doarXiv{2206.14297}

\bibitem[{{Cenko} {et~al.}(2012){Cenko}, {Krimm}, {Horesh}, {Rau}, {Frail},
  {Kennea}, {Levan}, {Holland}, {Butler}, {Quimby}, {Bloom}, {Filippenko},
  {Gal-Yam}, {Greiner}, {Kulkarni}, {Ofek}, {Olivares E.}, {Schady},
  {Silverman}, {Tanvir}, \& {Xu}}]{Cenko12}
{Cenko}, S.~B., {Krimm}, H.~A., {Horesh}, A., {et~al.} 2012, \apj, 753, 77,
  \dodoi{10.1088/0004-637X/753/1/77}

\bibitem[{{Chatterjee} {et~al.}(2020){Chatterjee}, {Younsi}, {Liska},
  {Tchekhovskoy}, {Markoff}, {Yoon}, {van Eijnatten}, {Hesp}, {Ingram}, \& {van
  der Klis}}]{Chatterjee20}
{Chatterjee}, K., {Younsi}, Z., {Liska}, M., {et~al.} 2020, \mnras, 499, 362,
  \dodoi{10.1093/mnras/staa2718}

\bibitem[{{Chevalier} \& {Li}(2000)}]{Chevalier}
{Chevalier}, R.~A., \& {Li}, Z.-Y. 2000, \apj, 536, 195, \dodoi{10.1086/308914}

\bibitem[{{Coughlin} \& {Begelman}(2014)}]{Coughlin&Begelman14}
{Coughlin}, E.~R., \& {Begelman}, M.~C. 2014, \apj, 781, 82,
  \dodoi{10.1088/0004-637X/781/2/82}

\bibitem[{{Coughlin} \& {Begelman}(2020)}]{Coughlin&Begelman20}
---. 2020, \mnras, 499, 3158, \dodoi{10.1093/mnras/staa3026}

\bibitem[{{Curd} \& {Narayan}(2019)}]{Curd&Narayan19}
{Curd}, B., \& {Narayan}, R. 2019, \mnras, 483, 565,
  \dodoi{10.1093/mnras/sty3134}

\bibitem[{{Dai} {et~al.}(2021){Dai}, {Lodato}, \& {Cheng}}]{Dai+21}
{Dai}, J.~L., {Lodato}, G., \& {Cheng}, R. 2021, \ssr, 217, 12,
  \dodoi{10.1007/s11214-020-00747-x}

\bibitem[{{Dai} {et~al.}(2018){Dai}, {McKinney}, {Roth}, {Ramirez-Ruiz}, \&
  {Miller}}]{Dai+18}
{Dai}, L., {McKinney}, J.~C., {Roth}, N., {Ramirez-Ruiz}, E., \& {Miller},
  M.~C. 2018, \apjl, 859, L20, \dodoi{10.3847/2041-8213/aab429}

\bibitem[{{Dexter} \& {Fragile}(2011)}]{Dexter&Fragile11}
{Dexter}, J., \& {Fragile}, P.~C. 2011, \apj, 730, 36,
  \dodoi{10.1088/0004-637X/730/1/36}

\bibitem[{{Eftekhari} {et~al.}(2018){Eftekhari}, {Berger}, {Zauderer},
  {Margutti}, \& {Alexander}}]{Eftekhari}
{Eftekhari}, T., {Berger}, E., {Zauderer}, B.~A., {Margutti}, R., \&
  {Alexander}, K.~D. 2018, \apj, 854, 86, \dodoi{10.3847/1538-4357/aaa8e0}

\bibitem[{{Evans} \& {Kochanek}(1989)}]{Evans&Kochanek89}
{Evans}, C.~R., \& {Kochanek}, C.~S. 1989, \apjl, 346, L13,
  \dodoi{10.1086/185567}

\bibitem[{{Fender} {et~al.}(2004){Fender}, {Belloni}, \& {Gallo}}]{Fender+04}
{Fender}, R.~P., {Belloni}, T.~M., \& {Gallo}, E. 2004, \mnras, 355, 1105,
  \dodoi{10.1111/j.1365-2966.2004.08384.x}

\bibitem[{{Foucart} \& {Lai}(2014)}]{Foucart&Lai14}
{Foucart}, F., \& {Lai}, D. 2014, \mnras, 445, 1731,
  \dodoi{10.1093/mnras/stu1869}

\bibitem[{{Fragile} {et~al.}(2007){Fragile}, {Blaes}, {Anninos}, \&
  {Salmonson}}]{Fragile+07}
{Fragile}, P.~C., {Blaes}, O.~M., {Anninos}, P., \& {Salmonson}, J.~D. 2007,
  \apj, 668, 417, \dodoi{10.1086/521092}

\bibitem[{{Franchini} {et~al.}(2016){Franchini}, {Lodato}, \&
  {Facchini}}]{Franchini+16}
{Franchini}, A., {Lodato}, G., \& {Facchini}, S. 2016, \mnras, 455, 1946,
  \dodoi{10.1093/mnras/stv2417}

\bibitem[{{Frank} {et~al.}(2002){Frank}, {King}, \& {Raine}}]{Frank+02}
{Frank}, J., {King}, A., \& {Raine}, D.~J. 2002, {Accretion Power in
  Astrophysics: Third Edition}

\bibitem[{{Generozov} {et~al.}(2017){Generozov}, {Mimica}, {Metzger}, {Stone},
  {Giannios}, \& {Aloy}}]{Generozov+17}
{Generozov}, A., {Mimica}, P., {Metzger}, B.~D., {et~al.} 2017, \mnras, 464,
  2481, \dodoi{10.1093/mnras/stw2439}

\bibitem[{{Giannios} \& {Metzger}(2011)}]{Giannios&Metzger11}
{Giannios}, D., \& {Metzger}, B.~D. 2011, \mnras, 416, 2102,
  \dodoi{10.1111/j.1365-2966.2011.19188.x}

\bibitem[{{Gomez} {et~al.}(2020){Gomez}, {Nicholl}, {Short}, {Margutti},
  {Alexander}, {Blanchard}, {Berger}, {Eftekhari}, {Schulze}, {Anderson},
  {Arcavi}, {Chornock}, {Cowperthwaite}, {Galbany}, {Herzog}, {Hiramatsu},
  {Hosseinzadeh}, {Laskar}, {M{\"u}ller Bravo}, {Patton}, \&
  {Terreran}}]{Gomez20}
{Gomez}, S., {Nicholl}, M., {Short}, P., {et~al.} 2020, \mnras, 497, 1925,
  \dodoi{10.1093/mnras/staa2099}

\bibitem[{{Granot} \& {Sari}(2002)}]{Granot&Sari02}
{Granot}, J., \& {Sari}, R. 2002, \apj, 568, 820, \dodoi{10.1086/338966}

\bibitem[{{Guillochon} \& {Ramirez-Ruiz}(2013)}]{Guillochon&RamirezRuiz13}
{Guillochon}, J., \& {Ramirez-Ruiz}, E. 2013, \apj, 767, 25,
  \dodoi{10.1088/0004-637X/767/1/25}

\bibitem[{{Horesh} {et~al.}(2021{\natexlab{a}}){Horesh}, {Cenko}, \&
  {Arcavi}}]{Horesh+21}
{Horesh}, A., {Cenko}, S.~B., \& {Arcavi}, I. 2021{\natexlab{a}}, Nature
  Astronomy, 5, 491, \dodoi{10.1038/s41550-021-01300-8}

\bibitem[{{Horesh} {et~al.}(2021{\natexlab{b}}){Horesh}, {Sfaradi}, {Fender},
  {Green}, {Williams}, \& {Bright}}]{Horesh+21b}
{Horesh}, A., {Sfaradi}, I., {Fender}, R., {et~al.} 2021{\natexlab{b}}, \apjl,
  920, L5, \dodoi{10.3847/2041-8213/ac25fe}

\bibitem[{{Horesh} {et~al.}(2018){Horesh}, {Sfaradi}, {Bright}, {Williams},
  {Fender}, {Titterington}, {Green}, {Perrott}, {Van Velzen}, \&
  {Gezari}}]{Horesh18}
{Horesh}, A., {Sfaradi}, I., {Bright}, J., {et~al.} 2018, The Astronomer's
  Telegram, 12271, 1

\bibitem[{{Kawamuro} {et~al.}(2016){Kawamuro}, {Ueda}, {Shidatsu}, {Hori},
  {Kawai}, {Negoro}, \& {Mihara}}]{Kawamuro16}
{Kawamuro}, T., {Ueda}, Y., {Shidatsu}, M., {et~al.} 2016, \pasj, 68, 58,
  \dodoi{10.1093/pasj/psw056}

\bibitem[{{Kelley} {et~al.}(2014){Kelley}, {Tchekhovskoy}, \&
  {Narayan}}]{Kelley+14}
{Kelley}, L.~Z., {Tchekhovskoy}, A., \& {Narayan}, R. 2014, \mnras, 445, 3919,
  \dodoi{10.1093/mnras/stu2041}

\bibitem[{{Lacy} {et~al.}(2020){Lacy}, {Baum}, {Chandler}, {Chatterjee},
  {Clarke}, {Deustua}, {English}, {Farnes}, {Gaensler}, {Gugliucci},
  {Hallinan}, {Kent}, {Kimball}, {Law}, {Lazio}, {Marvil}, {Mao}, {Medlin},
  {Mooley}, {Murphy}, {Myers}, {Osten}, {Richards}, {Rosolowsky}, {Rudnick},
  {Schinzel}, {Sivakoff}, {Sjouwerman}, {Taylor}, {White}, {Wrobel},
  {Andernach}, {Beasley}, {Berger}, {Bhatnager}, {Birkinshaw}, {Bower},
  {Brandt}, {Brown}, {Burke-Spolaor}, {Butler}, {Comerford}, {Demorest}, {Fu},
  {Giacintucci}, {Golap}, {G{\"u}th}, {Hales}, {Hiriart}, {Hodge}, {Horesh},
  {Ivezi{\'c}}, {Jarvis}, {Kamble}, {Kassim}, {Liu}, {Loinard}, {Lyons},
  {Masters}, {Mezcua}, {Moellenbrock}, {Mroczkowski}, {Nyland}, {O'Dea},
  {O'Sullivan}, {Peters}, {Radford}, {Rao}, {Robnett}, {Salcido}, {Shen},
  {Sobotka}, {Witz}, {Vaccari}, {van Weeren}, {Vargas}, {Williams}, \&
  {Yoon}}]{Lacy20}
{Lacy}, M., {Baum}, S.~A., {Chandler}, C.~J., {et~al.} 2020, \pasp, 132,
  035001, \dodoi{10.1088/1538-3873/ab63eb}

\bibitem[{{Larwood} \& {Papaloizou}(1997)}]{Larwood&Papaloizou97}
{Larwood}, J.~D., \& {Papaloizou}, J. C.~B. 1997, \mnras, 285, 288,
  \dodoi{10.1093/mnras/285.2.288}

\bibitem[{{Levan} {et~al.}(2011)}]{Levan+11}
{Levan}, A.~J., {et~al.} 2011, Science, 333, 199,
  \dodoi{10.1126/science.1207143}

\bibitem[{{Liska} {et~al.}(2018){Liska}, {Hesp}, {Tchekhovskoy}, {Ingram}, {van
  der Klis}, \& {Markoff}}]{Liska+18}
{Liska}, M., {Hesp}, C., {Tchekhovskoy}, A., {et~al.} 2018, \mnras, 474, L81,
  \dodoi{10.1093/mnrasl/slx174}

\bibitem[{{Lodato} {et~al.}(2009){Lodato}, {King}, \& {Pringle}}]{Lodato+09}
{Lodato}, G., {King}, A.~R., \& {Pringle}, J.~E. 2009, \mnras, 392, 332,
  \dodoi{10.1111/j.1365-2966.2008.14049.x}

\bibitem[{{Marti} {et~al.}(1994){Marti}, {Mueller}, \& {Ibanez}}]{Marti+94}
{Marti}, J.~M., {Mueller}, E., \& {Ibanez}, J.~M. 1994, \aap, 281, L9

\bibitem[{{Matsumoto} \& {Metzger}(2023)}]{Matsumoto&Metzger23}
{Matsumoto}, T., \& {Metzger}, B.~D. 2023, arXiv e-prints, arXiv:2301.11939,
  \dodoi{10.48550/arXiv.2301.11939}

\bibitem[{{Mattila} {et~al.}(2018){Mattila}, {P{\'e}rez-Torres}, {Efstathiou},
  {Mimica}, {Fraser}, {Kankare}, {Alberdi}, {Aloy}, {Heikkil{\"a}}, {Jonker},
  {Lundqvist}, {Mart{\'\i}-Vidal}, {Meikle}, {Romero-Ca{\~n}izales}, {Smartt},
  {Tsygankov}, {Varenius}, {Alonso-Herrero}, {Bondi}, {Fransson},
  {Herrero-Illana}, {Kangas}, {Kotak}, {Ram{\'\i}rez-Olivencia},
  {V{\"a}is{\"a}nen}, {Beswick}, {Clements}, \& {Greimel}}]{Mattila18}
{Mattila}, S., {P{\'e}rez-Torres}, M., {Efstathiou}, A., {et~al.} 2018,
  Science, 361, 482, \dodoi{10.1126/science.aao4669}

\bibitem[{{Matzner}(2003)}]{Matzner03}
{Matzner}, C.~D. 2003, \mnras, 345, 575,
  \dodoi{10.1046/j.1365-8711.2003.06969.x}

\bibitem[{{McKinney} {et~al.}(2013){McKinney}, {Tchekhovskoy}, \&
  {Blandford}}]{McKinney+13}
{McKinney}, J.~C., {Tchekhovskoy}, A., \& {Blandford}, R.~D. 2013, Science,
  339, 49, \dodoi{10.1126/science.1230811}

\bibitem[{{Metzger}(2022)}]{Metzger22}
{Metzger}, B.~D. 2022, \apjl, 937, L12, \dodoi{10.3847/2041-8213/ac90ba}

\bibitem[{{Metzger} {et~al.}(2012{\natexlab{a}}){Metzger}, {Giannios}, \&
  {Mimica}}]{Metzger+12}
{Metzger}, B.~D., {Giannios}, D., \& {Mimica}, P. 2012{\natexlab{a}}, \mnras,
  420, 3528, \dodoi{10.1111/j.1365-2966.2011.20273.x}

\bibitem[{{Metzger} {et~al.}(2008){Metzger}, {Piro}, \&
  {Quataert}}]{Metzger+08}
{Metzger}, B.~D., {Piro}, A.~L., \& {Quataert}, E. 2008, \mnras, 390, 781,
  \dodoi{10.1111/j.1365-2966.2008.13789.x}

\bibitem[{{Metzger} {et~al.}(2012{\natexlab{b}}){Metzger}, {Rafikov}, \&
  {Bochkarev}}]{Metzger+12b}
{Metzger}, B.~D., {Rafikov}, R.~R., \& {Bochkarev}, K.~V. 2012{\natexlab{b}},
  \mnras, 423, 505, \dodoi{10.1111/j.1365-2966.2012.20895.x}

\bibitem[{{Metzger} \& {Stone}(2016)}]{Metzger&Stone16}
{Metzger}, B.~D., \& {Stone}, N.~C. 2016, \mnras, 461, 948,
  \dodoi{10.1093/mnras/stw1394}

\bibitem[{{Mimica} {et~al.}(2015){Mimica}, {Giannios}, {Metzger}, \&
  {Aloy}}]{Mimica+15}
{Mimica}, P., {Giannios}, D., {Metzger}, B.~D., \& {Aloy}, M.~A. 2015, ArXiv
  e-prints.
\newblock \doarXiv{1501.00361}

\bibitem[{{Murphy} {et~al.}(2021){Murphy}, {Kaplan}, {Stewart}, {O'Brien},
  {Lenc}, {Pintaldi}, {Pritchard}, {Dobie}, {Fox}, {Leung}, {An}, {Bell},
  {Broderick}, {Chatterjee}, {Dai}, {d'Antonio}, {Doyle}, {Gaensler}, {Heald},
  {Horesh}, {Jones}, {McConnell}, {Moss}, {Raja}, {Ramsay}, {Ryder}, {Sadler},
  {Sivakoff}, {Wang}, {Wang}, {Wheatland}, {Whiting}, {Allison}, {Anderson},
  {Ball}, {Bannister}, {Bock}, {Bolton}, {Bunton}, {Chekkala}, {Chippendale},
  {Cooray}, {Gupta}, {Hayman}, {Jeganathan}, {Koribalski}, {Lee-Waddell},
  {Mahony}, {Marvil}, {McClure-Griffiths}, {Mirtschin}, {Ng}, {Pearce},
  {Phillips}, \& {Voronkov}}]{Murphy21}
{Murphy}, T., {Kaplan}, D.~L., {Stewart}, A.~J., {et~al.} 2021, \pasa, 38,
  e054, \dodoi{10.1017/pasa.2021.44}

\bibitem[{{Narayan} {et~al.}(2003){Narayan}, {Igumenshchev}, \&
  {Abramowicz}}]{Narayan+03}
{Narayan}, R., {Igumenshchev}, I.~V., \& {Abramowicz}, M.~A. 2003, \pasj, 55,
  L69, \dodoi{10.1093/pasj/55.6.L69}

\bibitem[{{Nelson} \& {Papaloizou}(2000)}]{Nelson&Papaloizou00}
{Nelson}, R.~P., \& {Papaloizou}, J. C.~B. 2000, \mnras, 315, 570,
  \dodoi{10.1046/j.1365-8711.2000.03478.x}

\bibitem[{{Ohsuga} \& {Mineshige}(2011)}]{Ohsuga&Mineshige11}
{Ohsuga}, K., \& {Mineshige}, S. 2011, \apj, 736, 2,
  \dodoi{10.1088/0004-637X/736/1/2}

\bibitem[{{Papaloizou} \& {Lin}(1995)}]{Papaloizou&Lin95}
{Papaloizou}, J.~C.~B., \& {Lin}, D.~N.~C. 1995, \apj, 438, 841,
  \dodoi{10.1086/175127}

\bibitem[{{Papaloizou} \& {Pringle}(1983)}]{Papaloizou&Pringle83}
{Papaloizou}, J.~C.~B., \& {Pringle}, J.~E. 1983, \mnras, 202, 1181,
  \dodoi{10.1093/mnras/202.4.1181}

\bibitem[{{Pasham} {et~al.}(2015){Pasham}, {Cenko}, {Levan}, {Bower}, {Horesh},
  {Brown}, {Dolan}, {Wiersema}, {Filippenko}, {Fruchter}, {Greiner},
  {Hounsell}, {O'Brien}, {Page}, {Rau}, \& {Tanvir}}]{Pasham+15}
{Pasham}, D.~R., {Cenko}, S.~B., {Levan}, A.~J., {et~al.} 2015, ArXiv e-prints.
\newblock \doarXiv{1502.01345}

\bibitem[{{Perlman} {et~al.}(2022){Perlman}, {Meyer}, {Wang}, {Yuan},
  {Henriksen}, {Irwin}, {Li}, {Wiegert}, {Li}, \& {Yang}}]{Perlman+22}
{Perlman}, E.~S., {Meyer}, E.~T., {Wang}, Q.~D., {et~al.} 2022, \apj, 925, 143,
  \dodoi{10.3847/1538-4357/ac3bba}

\bibitem[{{Phinney}(1989)}]{Phinney89}
{Phinney}, E.~S. 1989, in The Center of the Galaxy, ed. M.~{Morris}, Vol. 136,
  543

\bibitem[{{Polko} \& {McKinney}(2017)}]{Polko+17}
{Polko}, P., \& {McKinney}, J.~C. 2017, \mnras, 464, 2660,
  \dodoi{10.1093/mnras/stw1875}

\bibitem[{{Pringle}(1981)}]{Pringle81}
{Pringle}, J.~E. 1981, \araa, 19, 137,
  \dodoi{10.1146/annurev.aa.19.090181.001033}

\bibitem[{{Rees}(1988)}]{Rees88}
{Rees}, M.~J. 1988, \nat, 333, 523, \dodoi{10.1038/333523a0}

\bibitem[{{Reynolds}(2013)}]{Reynolds13}
{Reynolds}, C.~S. 2013, Classical and Quantum Gravity, 30, 244004,
  \dodoi{10.1088/0264-9381/30/24/244004}

\bibitem[{{Rhodes} {et~al.}(2023){Rhodes}, {Bright}, {Fender}, {Sfaradi},
  {Green}, {Horesh}, {Mooley}, {Pasham}, {Smartt}, {Titterington}, {van der
  Horst}, \& {Williams}}]{Rhodes}
{Rhodes}, L., {Bright}, J.~S., {Fender}, R., {et~al.} 2023, \mnras, 521, 389,
  \dodoi{10.1093/mnras/stad344}

\bibitem[{{Sadowski} \& {Narayan}(2015)}]{Sadowski&Narayan15}
{Sadowski}, A., \& {Narayan}, R. 2015, ArXiv e-prints.
\newblock \doarXiv{1503.00654}

\bibitem[{{Sadowski} \& {Narayan}(2016)}]{Sadowski&Narayan16}
---. 2016, \mnras, 456, 3929, \dodoi{10.1093/mnras/stv2941}

\bibitem[{{Saxton} {et~al.}(2017){Saxton}, {Read}, {Komossa}, {Lira},
  {Alexander}, \& {Wieringa}}]{Saxton17}
{Saxton}, R.~D., {Read}, A.~M., {Komossa}, S., {et~al.} 2017, \aap, 598, A29,
  \dodoi{10.1051/0004-6361/201629015}

\bibitem[{{Sfaradi} {et~al.}(2022){Sfaradi}, {Horesh}, {Fender}, {Green},
  {Williams}, {Bright}, \& {Schulze}}]{Sfaradi+22}
{Sfaradi}, I., {Horesh}, A., {Fender}, R., {et~al.} 2022, arXiv e-prints,
  arXiv:2202.00026.
\newblock \doarXiv{2202.00026}

\bibitem[{{Shakura} \& {Sunyaev}(1973)}]{Shakura&Sunyaev73}
{Shakura}, N.~I., \& {Sunyaev}, R.~A. 1973, \aap, 24, 337

\bibitem[{{Shen} \& {Matzner}(2014)}]{Shen&Matzner14}
{Shen}, R.-F., \& {Matzner}, C.~D. 2014, \apj, 784, 87,
  \dodoi{10.1088/0004-637X/784/2/87}

\bibitem[{{Somalwar} {et~al.}(2023){Somalwar}, {Ravi}, {Dong}, {Chen}, {Breen},
  {Chandra}, {Clarke}, {De}, {Gaensler}, {Hallinan}, {Laha}, {Law}, {Myers},
  {Parsotan}, {Peters}, \& {Polisensky}}]{Somalwar+23}
{Somalwar}, J.~J., {Ravi}, V., {Dong}, D.~Z., {et~al.} 2023, \apj, 945, 142,
  \dodoi{10.3847/1538-4357/acbafc}

\bibitem[{{Stone} \& {Loeb}(2012)}]{Stone&Loeb12}
{Stone}, N., \& {Loeb}, A. 2012, \prl, 108, 061302,
  \dodoi{10.1103/PhysRevLett.108.061302}

\bibitem[{{Stone} {et~al.}(2013){Stone}, {Sari}, \& {Loeb}}]{Stone+13}
{Stone}, N., {Sari}, R., \& {Loeb}, A. 2013, \mnras, 435, 1809,
  \dodoi{10.1093/mnras/stt1270}

\bibitem[{{Strubbe} \& {Quataert}(2009)}]{Strubbe&Quataert09}
{Strubbe}, L.~E., \& {Quataert}, E. 2009, \mnras, 400, 2070,
  \dodoi{10.1111/j.1365-2966.2009.15599.x}

\bibitem[{{Tchekhovskoy} {et~al.}(2014){Tchekhovskoy}, {Metzger}, {Giannios},
  \& {Kelley}}]{Tchekhovskoy+14}
{Tchekhovskoy}, A., {Metzger}, B.~D., {Giannios}, D., \& {Kelley}, L.~Z. 2014,
  \mnras, 437, 2744, \dodoi{10.1093/mnras/stt2085}

\bibitem[{{Tchekhovskoy} {et~al.}(2010{\natexlab{a}}){Tchekhovskoy}, {Narayan},
  \& {McKinney}}]{Tchekhovskoy+10}
{Tchekhovskoy}, A., {Narayan}, R., \& {McKinney}, J.~C. 2010{\natexlab{a}},
  \apj, 711, 50, \dodoi{10.1088/0004-637X/711/1/50}

\bibitem[{{Tchekhovskoy} {et~al.}(2010{\natexlab{b}}){Tchekhovskoy}, {Narayan},
  \& {McKinney}}]{Tchekhovskoy+10b}
---. 2010{\natexlab{b}}, \apj, 711, 50, \dodoi{10.1088/0004-637X/711/1/50}

\bibitem[{{Tchekhovskoy} {et~al.}(2011){Tchekhovskoy}, {Narayan}, \&
  {McKinney}}]{Tchekhovskoy+11}
---. 2011, Mon. Not. R. Astron. Soc., 418, L79,
  \dodoi{10.1111/j.1745-3933.2011.01147.x}

\bibitem[{{Teboul} \& {Shaviv}(2021)}]{Odelia1}
{Teboul}, O., \& {Shaviv}, N.~J. 2021, \mnras, 507, 5340,
  \dodoi{10.1093/mnras/stab2491}

\bibitem[{{Teboul} {et~al.}(2022){Teboul}, {Stone}, \& {Ostriker}}]{Odelia2}
{Teboul}, O., {Stone}, N.~C., \& {Ostriker}, J.~P. 2022, arXiv e-prints,
  arXiv:2211.05858, \dodoi{10.48550/arXiv.2211.05858}

\bibitem[{{van Velzen} {et~al.}(2013){van Velzen}, {Frail}, {K{\"o}rding}, \&
  {Falcke}}]{vanVelzen+13}
{van Velzen}, S., {Frail}, D.~A., {K{\"o}rding}, E., \& {Falcke}, H. 2013,
  \aap, 552, A5, \dodoi{10.1051/0004-6361/201220426}

\bibitem[{{van Velzen} {et~al.}(2016){van Velzen}, {Anderson}, {Stone},
  {Fraser}, {Wevers}, {Metzger}, {Jonker}, {van der Horst}, {Staley}, {Mendez},
  {Miller-Jones}, {Hodgkin}, {Campbell}, \& {Fender}}]{vanVelzen16}
{van Velzen}, S., {Anderson}, G.~E., {Stone}, N.~C., {et~al.} 2016, Science,
  351, 62, \dodoi{10.1126/science.aad1182}

\bibitem[{{Yao} {et~al.}(2023){Yao}, {Ravi}, {Gezari}, {van Velzen}, {Lu},
  {Schulze}, {Somalwar}, {Kulkarni}, {Hammerstein}, {Nicholl}, {Graham},
  {Perley}, {Cenko}, {Stein}, {Ricarte}, {Chadayammuri}, {Quataert}, {Bellm},
  {Bloom}, {Dekany}, {Drake}, {Groom}, {Mahabal}, {Prince}, {Riddle},
  {Rusholme}, {Sharma}, {Sollerman}, \& {Yan}}]{Yao+23}
{Yao}, Y., {Ravi}, V., {Gezari}, S., {et~al.} 2023, arXiv e-prints,
  arXiv:2303.06523, \dodoi{10.48550/arXiv.2303.06523}

\bibitem[{{Zauderer} {et~al.}(2013){Zauderer}, {Berger}, {Margutti}, {Pooley},
  {Sari}, {Soderberg}, {Brunthaler}, \& {Bietenholz}}]{Zauderer+13}
{Zauderer}, B.~A., {Berger}, E., {Margutti}, R., {et~al.} 2013, \apj, 767, 152,
  \dodoi{10.1088/0004-637X/767/2/152}

\bibitem[{{Zauderer} {et~al.}(2011)}]{Zauderer+11}
{Zauderer}, B.~A., {et~al.} 2011, \nat, 476, 425, \dodoi{10.1038/nature10366}

\end{thebibliography}
\bibliographystyle{aasjournal}





\end{document}